\documentclass[preprint,12pt]{elsarticle}

\usepackage{amssymb}
\usepackage{amsmath}

\usepackage{pgfplots}

\newcommand\bb{\boldsymbol{b}}
\newcommand\bB{\boldsymbol{B}}

\journal{Computer Physics Communications}

\begin{document}

\begin{frontmatter}



\title{TRIMEG-GKX: an electromagnetic gyrokinetic particle code with a Piecewise Field-Aligned Finite Element Method for Micro- and Macro-Instability Studies in Tokamak Core Plasmas}


\author{Zhixin Lu, Guo Meng, Roman Hatzky, Philipp Lauber, Matthias Hoelzl} 

\affiliation{organization={Max Planck Institute for Plasma Physics},
            addressline={Boltzmannstr. 2}, 
            city={ Garching},
            postcode={85748}, 
            country={Germany}}

\begin{abstract}
The features of the TRIMEG-GKX code are described with  emphasis on the exploration using novel/different schemes compared to other gyrokinetic codes, particularly the use of object-oriented programming, filter/buffer-free treatment, and a  high-order piecewise field-aligned finite element method. The TRIMEG-GKX code solves the electromagnetic gyrokinetic equation using the particle-in-cell scheme, taking into account multi-species effects and shear Alfv\'en physics. The mixed-variable/pullback scheme has been implemented to enable electromagnetic studies. This code is parallelized using particle decomposition and domain cloning among computing nodes, replacing traditional domain decomposition techniques. 

The applications to study the micro- and macro-instabilities are demonstrated, including the energetic-particle-driven Alfv\'en eigenmode, ion temperature gradient mode, and kinetic ballooning mode. Good performance is achieved in both ad hoc and experimentally reconstructed equilibria, such as those of the ASDEX Upgrade (AUG), Tokamak à configuration variable (TCV), and the Joint European Torus (JET).  Future studies of edge physics using the high-order $C^1$ finite element method for triangular meshes in the TRIMEG-C1 code will be built upon the same numerical methods.
\end{abstract}




\end{frontmatter}



\section{Introduction}
\label{sec:intro}
Since the early development of the gyrokinetic simulation model \cite{lee1983gyrokinetic}, important nonlinear physics phenomena have been identified in the past decades, such as the generation of the zonal flows \cite{lin1998turbulent} and the turbulence impact on the edge heat fluxes to the divertors \cite{chang2017fast}. To enhance the capability of the modern gyrokinetic particle codes, various advanced numerical and physical schemes have been developed, such as the noise reduction scheme \cite{hatzky2019reduction} and different electromagnetic schemes such as the iterative $p_\|$ scheme \cite{chen2007electromagnetic}, the mixed variable/pullback scheme \cite{mishchenko2019pullback}, the implicit scheme \cite{lu2021development,sturdevant2021verification}, the GK-E\&B model and its implementation \cite{chen2021gyrokinetic,rosen2022and} and the conservative scheme \cite{bao2018conservative}. Advanced computational schemes and comprehensive physics models are key ingredients in modern gyrokinetic codes for successful prediction and interpretation of experimental results and underlying physics \cite{lanti2020orb5,kleiber2024euterpe,taimourzadeh2019verification}. New frameworks and methods have also been developed for high-fidelity particle simulations using the geometric particle-in-cell method \cite{kraus2017gempic,meng2025geometric}. 

The gyrokinetic codes TRIMEG-GKX based on structured meshes \cite{lu2021development,lu2023full,lu2025piecewise,meng2020effects} and TRIMEG-C0/C1 based on triangular unstructured meshes \cite{lu2019development,lana2023neoclassical,lu2024gyrokinetic} have been developed to explore various new methods for physics studies.  Since the early development using the unstructured meshes \cite{lu2019development}, physics models and numerical scheme, such as the implicit scheme \cite{lu2021development}, the full-$f$ mixed-variable/pullback scheme \cite{lu2023full}, and the piecewise field-aligned finite element method \cite{lu2025piecewise}, have been developed, demonstrating the capability of the various physics and numerical methods in gyrokinetic studies. Specifically, it is shown that using a high-order finite element method, the capability of the gyrokinetic code can be enhanced to capture the features of the instabilities using a lower grid resolution \cite{lu2024gyrokinetic}. In this work, the recent development of the TRIMEG-GKX code (referred to as GKX in the following) is demonstrated. Compared with other gyrokinetic particle codes, the features of the GKX code are as follows:
\begin{enumerate}
    \item Object-oriented programming principles are adopted, enabling the inclusion of multiple particle species and different physics models, such as the electrostatic or electromagnetic gyrokinetic models.
    \item The piecewise field-aligned finite element method (PFAFEM) is used with the cubic B-splines applied in the radial, poloidal, and toroidal directions. 
    \item The utilization of the Fourier filter is minimized, especially in the 3D electromagnetic model. Instead, a rigorous treatment of the strong form in the gyrocenter equation of motion is adopted.  
    \item The cache performance is optimized for particle pushers differently than the sorting scheme frequently used by other gyrokinetic particle codes. 
    \item The particle decomposition is adopted but the domain decomposition is replaced by the domain cloning among computing nodes due to the following reasons: (1) Shared memory is utilized through the MPI-3 standard, reducing total memory consumption on each computing node; (2) PFAFEM is applied, allowing to reduce the number of grid points for the 3D grid in one dimension by approximately one order of magnitude; (3) Cubic B-splines are adopted and thus four basis functions in the toroidal direction are involved.  
\end{enumerate}

The remainder of this article is organized as follows. Section \ref{sec:model} presents the models and equations implemented in the GKX code. Section \ref{sec:numeric_scheme} describes the numerical schemes. In Section \ref{sec:results}, simulation results are shown for various cases, including  ad hoc equilibria and realistic experimental configurations. Finally, conclusions and perspectives for future work are discussed in Section~\ref{sec:conclusion}.

\section{Models and equations}
\label{sec:model}
The physics model employed in this work is closely related to the early work on the pullback scheme \cite{mishchenko2014pullback} and our previous development of the mixed full $f$-$\delta f$ scheme \cite{lu2023full}, as well as the references therein \cite{hatzky2019reduction,mishchenko2019pullback,lanti2020orb5}. The focus of this article is on the traditional $\delta f$ model.

\subsection{Physics equations using mixed variables}
The mixed variable is defined as follows.
The parallel component $\delta A_\|$ of the perturbed magnetic potential is decomposed into a symplectic part and a Hamiltonian part \cite{mishchenko2014pullback},
\begin{equation}
\label{eq:AsAh}
    \delta A_\| =\delta A_\|^{\rm{s}} + \delta A_\|^{\rm{h}} \;\;, 
\end{equation}
where the symplectic part $\delta A_\|^{\rm{s}}$ is chosen to satisfy ideal Ohm's law involving the electrostatic scalar potential $\delta\Phi$ as follows,
\begin{equation}
    \partial_t\delta A_\|^{\rm{s}}+\partial_\|\delta\Phi = 0\;\;,
\label{eq:ohm_law0}
\end{equation}
where the parallel derivative is defined as $\partial_\|=\boldsymbol{b}\cdot\nabla$, $\boldsymbol{b}=\boldsymbol{B}/B$, $\boldsymbol{B}$ is the equilibrium magnetic field. 
The shifted parallel velocity coordinate of the gyrocenter $u_\|$ is defined as 
\begin{equation}
    u_\|=v_\|+\frac{q_s}{m_s}\langle\delta A_\|^{\rm{h}}\rangle\;\;,
\end{equation}
where $v_\|$ is the parallel velocity, $q_s$ and $m_s$ are the charge and mass of species $s$, respectively, the subscript ``$s$'' represents the different particle species, and $\langle\ldots\rangle$ indicates the gyro average.

The gyrocenter equations of motion are consistent with previous work \cite{mishchenko2019pullback,hatzky2019reduction,lanti2020orb5,mishchenko2023global,kleiber2024euterpe}, {
\begin{eqnarray}
 \dot{\boldsymbol R}_0 
  &=& u_\| {\boldsymbol b}^*_0 + \frac{m\mu}{qB^*_\|} {\boldsymbol b}\times\nabla B \;\;, 
  \\
  \dot u_{\|,0}
  &=& -\mu {\boldsymbol b}^*_0\cdot \nabla B \;\;,
  \\
  \delta\dot{\boldsymbol R}
  &=& \frac{{\boldsymbol b}}{B^*_\|}\times \nabla \langle \delta\Phi -u_\| \delta A_\|\rangle 
  -\frac{q_s}{m_s}\langle\delta A^{\rm h}_\|\rangle {\boldsymbol b}^*\;\;, 
  \\
  \delta \dot u_\|
  &=&  -\frac{q_s}{m_s} \left({\boldsymbol b}^*\cdot\nabla\langle\delta\Phi-u_\|\delta A^{\rm{h}}_\|\rangle +\partial_t\langle\delta A_\|^{\rm{s}}\rangle \right) \nonumber\\
  &&-\frac{\mu}{B^*_\|}{\boldsymbol b}\times\nabla B\cdot\nabla\langle\delta A_\|^{\rm{s}}\rangle \;\;,  
\end{eqnarray}
where the magnetic moment $\mu=v_\perp^2/(2B)$, ${\boldsymbol b}^*={\boldsymbol b}_0^*+\nabla\langle\delta A_\|^{\rm s}\rangle\times{\boldsymbol b}/B_\|^*$, $\langle\ldots\rangle$ denotes the gyro average, ${\boldsymbol b}^*_0={\boldsymbol b}+(m_s/q_s)u_\|\nabla\times{\boldsymbol b}/B_\|^*$, ${\boldsymbol b}={\boldsymbol B}/B$, $B_\|^*=B+(m_s/q_s)u_\|{\boldsymbol b}\cdot(\nabla\times{\boldsymbol b})$. Note that in our previous work \cite{lu2023full}, $v_\|$ is adopted on the right-hand side, and  the term $-(q_s/m_s)\langle\delta A_\|^{\rm{h}}\rangle{\boldsymbol b}^*$ in $\dot{\boldsymbol R}$ is thus taken into account in $\dot{\boldsymbol R}_0$. The expression of the gyrocenter equation of motion in $(r,\phi,\theta)$ coordinates is listed in  \ref{gcmotion}.

\subsection{Field equations}
\label{subsec:field_equations}
The linearized quasi-neutrality equation in the long-wavelength approximation is as follows, 
\begin{equation}
\label{eq:poisson0}
    -\nabla\cdot\left( \sum_s\frac{q_s n_{0s}}{B\omega_{{\rm c}s}} \nabla_\perp\delta\Phi \right) = \sum_s q_s \delta n_{s,v} \;\;,
\end{equation}
where the gyrocenter density $\delta n_{s,v}$ is calculated using $\delta f_s({\boldsymbol R},v_\|,\mu)$ (indicated as $\delta f_{s,v}$), namely, $\delta n_{s,v}({\boldsymbol{x} })=\int {\rm d}^6 z\delta f_{s,v}\delta(\boldsymbol{R}  + \boldsymbol{\rho} - \boldsymbol{x} )$. Here, $\boldsymbol{x} $ and $\boldsymbol{R} $ denote the particle and gyrocenter coordinates, respectively, and $\boldsymbol{\rho}$ represents the Larmor radius.
In Eq.~\eqref{eq:poisson0}, $\omega_{{\rm c}s}$ is the cyclotron frequency of species $s$, and in this work, we ignore the perturbed electron polarization density on the left-hand side. 
When the $\delta f$ scheme is adopted, $\delta f_{s,v}$ is obtained from $\delta f_{s,u}$ as follows, with the linear approximation of the pullback scheme,
\begin{eqnarray}
    & \delta f_{s,v} = \delta f_{s,u} +  \frac{q_s\left\langle\delta A^{\rm{h}}_{\|} \right\rangle}{m_s}\frac{\partial f_{0s}}{\partial v_\|}
    \xrightarrow[f_{0s}=f_{\rm{M}}]{\text{Maxwellian}}
     \delta f_{s,u} -  \frac{ v_\|}{T_s}  q_s\left\langle\delta A^{\rm{h}}_{\|} \right\rangle f_{0s} \;\;
\end{eqnarray}
produced from the general form of the nonlinear pullback scheme \cite{hatzky2019reduction},
\begin{eqnarray}
    & f_{s,v} (v_\|) = f_{s,u} \left(v_\|+\frac{q_s}{m_s}\langle\delta A_\|^{\rm h}\rangle\right)\;\;.
\end{eqnarray}

Amp\`ere's law in $v_\|$ space is given by 
\begin{equation}
    -\nabla^2_\perp\delta A_\| = \mu_0 \delta j_{\|,v} \;\;,
\end{equation}
where $\delta j_{\|,v}({\boldsymbol{x} })=\sum_s q_s \int {\rm d}^6 z\delta f_{s,v}\delta(\boldsymbol{R}  + \boldsymbol{\rho} - \boldsymbol{x} )v_\|$.

For the $\delta f$ model, using the mixed variables and assuming a Maxwell distribution, we have
\begin{eqnarray}
    \delta j_{\|,v}
    &\equiv&\sum_s q_s\int \mathrm{d}z^6 v_\| \delta f_{s,v}(v_\|)\delta(\boldsymbol{R}  + \boldsymbol{\rho} - \boldsymbol{x} )  \nonumber \\
    &=&\sum_s q_s\int \mathrm{d}z^6   v_\| \left(\delta f_{s,u}(u_\|) -\frac{v_\| q_s\langle \delta A_\|^{\rm{h}}\rangle }{T_s}f_{0s}\right) \delta(\boldsymbol{R}  + \boldsymbol{\rho} - \boldsymbol{x} )  \;\;.
\end{eqnarray}
Then we can write Amp\`ere's law as {
\begin{eqnarray}
\label{eq:ampere_mv_deltaf_exact}
    -\nabla^2_\perp\delta A_{\|}^{\rm{h}}
    +\sum_s\mu_0\frac{q_s^2}{T_s}\int \mathrm{d}z^6 v_\|^2 f_{0s} \langle \delta A_{\|}^{\rm{h}} \rangle \delta(\boldsymbol{R}  + \boldsymbol{\rho} - \boldsymbol{x} ) \nonumber\\
    =\nabla^2_\perp\delta A_{\|}^{\rm{s}} 
    + \mu_0\sum_s q_s  \int \mathrm{d}z^6  v_\| \delta f_{s,u}(u_\|)\delta(\boldsymbol{R}  + \boldsymbol{\rho} - \boldsymbol{x} )   \;\;.
\end{eqnarray}}
The integral on the left-hand side can be obtained analytically, yielding
\begin{eqnarray}
\label{eq:ampere_mv_deltaf}
    -\nabla^2_\perp\delta A_{\|}^{\rm{h}}
    &+&\sum_s\frac{1}{d_{s}^2}\overline{\langle\delta A_{\|}^{\rm{h}}\rangle  }
    =\nabla^2_\perp\delta A_{\|}^{\rm{s}} 
     \nonumber \\
     &+&\mu_0\sum_s q_s  \int \mathrm{d}z^6  v_\| \delta f_{s,u}(u_\|)  \delta(\boldsymbol{R}  + \boldsymbol{\rho} - \boldsymbol{x} ) \;\;, 
\end{eqnarray}
\begin{eqnarray}
\label{eq:dAh_from_f0}
    \overline{\langle\delta A_{\|}^{\rm{h}}\rangle  }
    &\equiv&\frac{2}{n_{s0}v_{ts}^2}\int \mathrm{d}z^6 v_\|^2 f_{0s} \langle \delta A_{\|}^{\rm{h}} \rangle \delta(\boldsymbol{R}  + \boldsymbol{\rho} - \boldsymbol{x} ) \,\,,
\end{eqnarray}
where $v_{\rm{t}s}=\sqrt{2T_s/m_s}$, $d_{s}$ is the skin depth of species $s$ defined as $d_{s}^2=c^2/\omega_{p,s}^2=m_s/(\mu_0q_s^2n_{0s})$ and the integral in Eq.~(\ref{eq:dAh_from_f0}) is kept without analytical reduction in order to capture the numerical or physics deviation of $f_0$ away from the Maxwellian distribution. 

For the full $f$ model, the perturbed current is represented by the full $f$,
\begin{eqnarray}
     j_{\|,v}&=&\sum_s q_s\int \mathrm{d}v^3 v_\| f_{s,v}  \nonumber \\
    &=&  \sum_sq_s\int \mathrm{d}z^6 \left[ u_\|-\frac{q_s}{m_s}\langle\delta A_\|^{\rm{h}}\rangle\right] f_{s,v} \delta(\boldsymbol{R}  + \boldsymbol{\rho} - \boldsymbol{x} )\;\;. 
\end{eqnarray} 
Amp\`ere's law yields {
\begin{eqnarray}
    -\nabla^2_\perp\delta A_{\|}^{\rm{h}}
    +\mu_0\sum_s\frac{q_s^2}{m_s}\int \mathrm{d}z^6 f_{s,v}\langle \delta A_{\|}^{\rm{h}} \rangle\delta(\boldsymbol{R}  + \boldsymbol{\rho} - \boldsymbol{x} )  \nonumber \\
    =\nabla^2_\perp\delta A_{\|}^{\rm{s}} + \mu_0\sum_s q_s  \int \mathrm{d}z^6  u_\| f_{s,v} \delta(\boldsymbol{R}  + \boldsymbol{\rho} - \boldsymbol{x} ) \;\;.
\end{eqnarray}}
The corresponding analytical limit gives a similar form as Eq.~(\ref{eq:ampere_mv_deltaf}) except the replacement of $\delta f_{s,u}(u_\|)$ with $f_{s,v}(v_\|)$ and the definition of $\overline{\langle\delta A_{\|}^{\rm{h}}\rangle  }$,
\begin{eqnarray}
    -\nabla^2_\perp\delta A_{\|}^{\rm{h}}
    +\sum_s\frac{1}{d_{s}^2}\overline{\langle\delta A_{\|}^{\rm{h}}\rangle  }& =&
    \nabla^2_\perp\delta A_{\|}^{\rm{s}} 
     \nonumber \\
    &+&\mu_0\sum_s q_s  \int \mathrm{d}z^6  v_\| f_{s,v}(v_\|)  \delta(\boldsymbol{R}  + \boldsymbol{\rho} - \boldsymbol{x} )  \;\;, 
\end{eqnarray}
with 
\begin{eqnarray}
    \overline{\langle\delta A_{\|}^{\rm{h}}\rangle  }
    &\equiv&
    \frac{1}{n_{s0}}\int \mathrm{d}z^6  f_{s,v} (v_\parallel) \langle \delta A_{\|}^{\rm{h}} \rangle \delta(\boldsymbol{R}  + \boldsymbol{\rho} - \boldsymbol{x} ) \;\;.
\end{eqnarray}

For both the $\delta f$ model and the full $f$ model, using an iterative scheme, the asymptotic solution is expressed as follows,
\begin{equation}
    \delta A^{\rm{h}}_{\|}=\sum_{I=0}^\infty\delta A^{\rm{h}}_{\|,I}\;\;,
\end{equation}
where $|\delta A^{\rm{h}}_{\|,I+1}/\delta A^{\rm{h}}_{\|,I}|\ll1$ is assured by the fact that the analytical skin depth term (the second term on the left-hand side of Eq.~(\ref{eq:ampere_mv_deltaf})) is close to the exact one (the second term on the right-hand side of Eq.~(\ref{eq:ampere_mv_deltaf})). 
Amp\`ere's law is solved order by order,
\begin{eqnarray}
\label{eq:ampere_h0}
    \left(\nabla^2_\perp-\sum_s\frac{1}{d_{s}^2}\right)\delta A_{\|,0}^{\rm{h}} 
    = -\nabla^2_\perp\delta A_{\|}^{\rm{s}} - \mu_0 \delta j_{\|} \;\;, \\
\label{eq:ampere_iterative}
    \left(\nabla^2_\perp-\sum_s\frac{1}{d_{s}^2}\right)\delta A_{\|,I}^{\rm{h}} 
    =-\sum_s\frac{1}{d_{s}^2}\delta A^{\rm{h}}_{\|,I-1} 
    + \sum_s\frac{1}{d_{s}^2} \overline{\langle\delta A_{\|,I-1}^{\rm{h}}\rangle}\;\;, \\
    \overline{\langle\delta A_{\|,I-1}^{\rm{h}}\rangle}
    =\frac{2}{n_0 v_{\rm{t}s}^2}\int \mathrm{d}z^6 v_\|^2 f_{0s}  \langle\delta A^{\rm{h}}_{\|,I-1} \rangle\delta(\boldsymbol{R}  + \boldsymbol{\rho} - \boldsymbol{x} )  \;\;\text{ for $\delta f$ model}\;\;, \\
\label{eq:A2ndavg_fullf}
    \overline{\langle\delta A_{\|,I-1}^{\rm{h}}\rangle}
    =\frac{1}{n_0}\int \mathrm{d}z^6 f_{s,v}  \langle\delta A^{\rm{h}}_{\|,I-1} \rangle \delta(\boldsymbol{R}  + \boldsymbol{\rho} - \boldsymbol{x} ) \;\;\text{ for full $f$ model \;\;,}
\end{eqnarray}
where  $I=1,2,3,\ldots$. 
For a Maxwellian distribution, when the finite Larmor radius (FLR) effect is neglected---as is appropriate for electrons, the primary contributors to the skin depth term---we have $2/(n_0 v_{\rm{t}s}^2)\int \mathrm{d}v^3 v_\|^2 f_0=1$ and $(1/n_0)\int \mathrm{d}v^3  f_0=1$.
Under these conditions, the analytical expression for the electron skin depth term closely approximates the exact value. As a result, the iterative solver is expected to converge well, enabling efficient and accurate computation of the skin depth contribution.


\subsection{Pullback scheme for mitigating the cancellation problem}
A more detailed description of the pullback scheme for the $\delta f$ method can be found in the previous work \cite{mishchenko2019pullback}, and the pullback scheme for full $f$ and mixed full $f$-$\delta f$ methods \cite{lu2023full}. As a brief review, the equations for $\delta f$ are listed as follows.
\begin{align}
\label{eq:pullback_A}
    & \delta A^{\rm{s}}_{\|,\rm{new}} = \delta A^{\rm{s}}_{\|,\rm{old}} + \delta A^{\rm{h}}_{\|,\rm{old}}  \;\;, \\
\label{eq:pullback_v}
    & u_{\|,\rm{new}} = u_{\|,\rm{old}} - \frac{q_s}{m_s} \left\langle\delta A^{\rm{h}}_{\|,\rm{old}} \right\rangle  \;\;, \\
\label{eq:pullback_df}
    & \delta f_{\rm{new}} = \delta f_{\rm{old}} +  \frac{q_s\left\langle\delta A^{\rm{h}}_{\|,\rm{old}} \right\rangle}{m_s}\frac{\partial f_{0s}}{\partial v_\|}
    \xrightarrow[f_{0s}=f_{M}]{\text{Maxwellian}}
     \delta f_{\rm{old}} -  \frac{ 2v_\|}{v_{\rm{t}s}^2}  \frac{q_s\left\langle\delta A^{\rm{h}}_{\|,\rm{old}} \right\rangle}{m_s} f_{0s} \;\;,
\end{align}
where variables with subscripts ``new'' and ``old'' refer to those after and before the pullback transformation, Eq.~(\ref{eq:pullback_df}) is the linearized equation for $\delta f$ pullback, which is from the general equation of the transformation for the distribution function 
\begin{align}
    f_{\rm{old}} (u_{\| \rm{old}}) = f_{\rm{new}} \left(u_{\| \rm{new}} =u_{\| \rm{old}}- \frac{q_s}{m_s} \left\langle\delta A^{\rm{h}}_{\|,\rm{old}} \right\rangle \right) \;\;. 
\end{align}
For the full $f$ scheme, only Eqs.~(\ref{eq:pullback_A}) and (\ref{eq:pullback_v}) are needed. 

\subsection{Discretization of distribution function}
Using the particle simulation scheme, $N$ markers are used with a given distribution,
\begin{align}
    g(z,t)\approx \sum_{p=1}^{N} \frac{\delta[z_p-z_p(t)]}{J_z}\;\;,
\end{align}
where $z$ is the phase space coordinate, $\delta$ is the Dirac delta function, $J_z$ is the corresponding Jacobian and $z=({\boldsymbol R},v_\|,\mu)$ is the phase space coordinate adopted in this work, $\mu$ is the magnetic moment, $v_\|$ is the parallel velocity, ${\boldsymbol R}$ is the real space coordinate. The total distribution of particles is represented by the markers,
\begin{align}
    f(z,t)=C_{\rm{g2f}} P_{\rm{tot}}(z,t)g(z,t)
    \approx C_{\rm{g2f}} \sum_{p=1}^{N} p_{p,\rm{tot}}(t) \frac{\delta[z_p-z_p(t)]}{J_z}\;\;,
\end{align}
where the constant $C_{\rm{g2f}}\equiv N_f/N_g$, $N_{f/g}$ is the number of particles/markers, and $g$ and $f$ refer to the markers and physical particles, respectively. 
For collisionless plasmas, 
\begin{align}
    \frac{{\rm d}g(z,t)}{{\rm d}t}=0 \;\;, \;\;
    \frac{{\rm d}f(z,t)}{{\rm d}t}=0 \;\;. 
\end{align}
For each marker, 
\begin{align}
    p_{p,\rm{tot}}(t)=\frac{1}{C_{\rm{g2f}}}\frac{f(z_p,t)}{g(z_p,t)}
    =\frac{N_g}{N_f}\frac{n_f}{n_g} \frac{f_v(z_p,t)}{g_v(z_p,t)}
    =\frac{n_f}{\langle n_f\rangle_V}\frac{\langle n_g \rangle_V}{n_g} \frac{f_v(z_p,t)}{g_v(z_p,t)} \;\;,
\end{align}
where  $n_f$ is the density profile and $f_v$ is the distribution in velocity space, namely, the particle distribution function $f=n_f({\boldsymbol R})f_v(v_\parallel,\mu)$, $\langle \ldots\rangle_V$ indicates the volume average. There are different choices of the marker distribution functions as discussed previously \cite{hatzky2019reduction,lanti2020orb5}. In this work, the markers are randomly initialized, uniformly distributed in the toroidal direction and in the $(R, Z)$ plane, while their velocity space distribution is identical to that of the physical particles, which leads to 
\begin{align}\label{eq:ptot_general}
    p_{p,\rm{tot}}(z,t) =\frac{\phi_{\rm wid}SR}{V_{\rm tot}}\;\;,
\end{align}
where $\phi_{\rm wid}$ is the width of the simulation domain in the toroidal direction, $S$ is the area of the poloidal cross-section, $V_{\rm tot}$ is the total volume. Equation~(\ref{eq:ptot_general}) is reduced to $p_{p,\rm{tot}}(z,t)={n_f}{R}/({\langle n_f\rangle_V}{R_0})$ for the tokamak equilibrium with concentric circular flux surfaces.

For the $\delta f$ model, the total distribution function is decomposed into the background and perturbed parts, $f(z,t)=f_0(z,t)+\delta f(z,t)$. The background part can be chosen as a time-independent one, i.e., $f_0(z,t)=f_0(z)$, and a typical choice is the local Maxwellian distribution. The background and perturbed distribution functions are represented by the markers as follows,
\begin{align}
    f_0(z,t)=P(z,t)g(z,t)\approx \sum_{p=1}^{N} p_p(t) \frac{\delta[z_p-z_p(t)]}{J_z}\;\;, \\
    \delta f(z,t)=W(z,t)g(z,t)\approx \sum_{p=1}^{N} w_p(t) \frac{\delta[z_p-z_p(t)]}{J_z}\;\;,
\end{align}
where $p_{p}(t)=f_0(z_p,t)/g(z_p,t)=P(z_p,t)$ and $w_{p}(t)=\delta f(z_p,t)/g(z_p,t)=W(z_p,t)$ are time-varying variables. 

\subsection{The evolution of the distribution function}
The evolution of the distribution function is represented by that of the marker weight $w_p$ and $p_p$.
The evolution equations are readily obtained \cite{lanti2020orb5},
\begin{align}
    &\dot w_p(t)  = - p_p(t) \frac{{\rm d}}{{\rm d}t} \left[\ln f_0(z_p(t)) \right] \;\;, \\
    &\dot p_p(t) =   p_p(t) \frac{{\rm d}}{{\rm d}t} \left[\ln f_0(z_p(t)) \right] \;\;, \\
    &\frac{{\rm d}}{{\rm d}t} = \frac{\partial}{\partial t} + \boldsymbol{\dot R}\cdot\nabla + \dot v_\|\frac{\partial}{\partial v_\|} \;\;,
\end{align}
where  the magnetic moment is assumed to be constant for the last equation, i.e., $\dot\mu= 0$. Generally, the gyrocenter equation of motion can be decomposed into the equilibrium part related to the equilibrium magnetic field and the perturbed part related to the perturbed field 
\begin{eqnarray}
  {\boldsymbol{\dot R}} ={\boldsymbol{\dot R}}_0 +\delta {\boldsymbol{\dot R}} \;\;, \\
  \dot v_\| = \dot v_{\|,0}+\delta \dot{v} _\| \;\;.
\end{eqnarray}
The time derivative is defined as 
\begin{eqnarray}
    \frac{\rm d}{{\rm d}t}&=&\left.\frac{{\rm d}}{{\rm d}t}\right|_0 +\left.\frac{{\rm d}}{{\rm d}t}\right|_1 \;\;, \\
    \left.\frac{{\rm d}}{{\rm d}t}\right|_0&=& \frac{\partial}{\partial t} + {\boldsymbol{\dot R}}_0\cdot\nabla + \dot v_{\|,0} \frac{\partial}{\partial v_\|}  \;\;,\\
     \left.\frac{{\rm d}}{{\rm d}t}\right|_1&=&\delta {\boldsymbol{\dot R}} \cdot\nabla + \delta \dot v_{\|} \frac{\partial}{\partial v_\|} \;\;.
\end{eqnarray}
For the equilibrium distribution function,
\begin{eqnarray}
 \left.\frac{{\rm d}}{{\rm d}t}\right|_0f_0=0\;\;,
\end{eqnarray}
and thus
\begin{eqnarray}
 \frac{{\rm d}}{{\rm d}t} f_0 = \left.\frac{{\rm d}}{{\rm d}t}\right|_1f_0 \;\;,
\end{eqnarray}
where $f_0$ is chosen as a steady state solution ($\partial f_0/\partial t=0$).
In this work, the local Maxwell distribution is chosen ($f_0=f_{\rm{M}}$), 
\begin{eqnarray}
 f_{\rm{M}}=\frac{n_0}{(2T/m)^{3/2}\pi^{3/2}} \exp\left(-\frac{mv_\|^2}{2T}-\frac{m\mu B}{T}\right) \;\;,
\end{eqnarray}
with $\int f_{\rm{M}} J_v dv_\|d\mu=n_0,\;\; J_v=2\pi B$. And thus
\begin{eqnarray}
 \frac{{\rm d}}{{\rm d}t}\ln f_{\rm{M}} = \delta {\bf\dot R}\cdot\left[
 {\vec\kappa}_n + \left(\frac{mv_\|^2}{2T}+\frac{m\mu B}{T}-\frac{3}{2}\right)\vec\kappa_T
 -\frac{m\mu B}{T}\vec\kappa_B
 \right]
 - \delta \dot v_{\|} \frac{mv_\|}{T} \;\;, \nonumber\\
\end{eqnarray}
where $\vec\kappa_{n,T,B}\equiv \nabla\ln \{n,T,B\}$. 
Note that for the local Maxwell distribution, the following approximation has been made in the  $\delta f$ scheme in this work: ${\rm d}f_{\rm{M}}/{\rm d}t|_0=0$ is assumed to eliminate the neoclassical drive.

\section{Numerical schemes}
\label{sec:numeric_scheme}
\subsection{Code structure based on object-oriented programming}
The GKX code is based on structured meshes to study the core plasmas in tokamaks \cite{lu2021development,lu2023full}. It is written in Fortran. Object Oriented Programming (OOP) principles are adopted with a similar structure to the TRIMEG-C0/C1 code based on the unstructured meshes \cite{lu2019development,lu2024gyrokinetic}. The gyrokinetic field-particle system is decomposed into different classes, namely, equilibrium, particle, field, solver, and B-spline classes. The application of the gyrokinetic field-particle classes is constructed by other basic classes. The kernel of the Fortran code is about 20,000 lines. The PETSc library is adopted to solve the linear field equations using the KSP solver. The shared memory feature in the MPI3 standard is utilized to store the 3D field, ensuring efficient memory usage. The equilibrium variables are represented using the B-splines \cite{williams}. The FEM is implemented using cubic B-splines with the details provided in our previous work \cite{lu2023full}. 

The structure of the code is demonstrated in Fig.~\ref{fig:uml}. The three physics classes are the \emph{equilibrium}, \emph{field}, and \emph{particle} classes. Other classes, such as the \emph{solver}, \emph{spline}, and \emph{gkem2sp} classes, are designed for numerical schemes and physics applications. The brief description of these classes is as follows. 
\begin{enumerate}
    \item The \emph{equilibrium} class treats the analytical and numerical (EQDSK) equilibrium and provides equilibrium data to other classes.
    \item The \emph{particle} class solves the gyrocenter equations of motion, treats the weight equation, and  performs the projection (``scattering'').
    \item The \emph{field} class defines the field equations and uses the \emph{spline} and \emph{solver} classes to solve the field equations. It also provides the fields to the particles by interpolation using the \emph{spline} class (``gathering''). 
    \item The \emph{particle coordinate} class is used in the time integrator to store the intermediate information. 
    \item The \emph{solver} class deals with the storage of the matrix and the linear system's solving process. 
    \item The \emph{3D cubic spline} class provides the variables and algorithms related to the 3D B-spline finite element method. 
    \item The \emph{gyrokinetic electromagnetic} class owns an instance of the \emph{equilibrium} class, an instance of the \emph{field} class, and an instance array of the \emph{particle} class (so that multiple species are included).
\end{enumerate} 

\begin{figure}[t]
\centering
\includegraphics[width=0.98\textwidth]{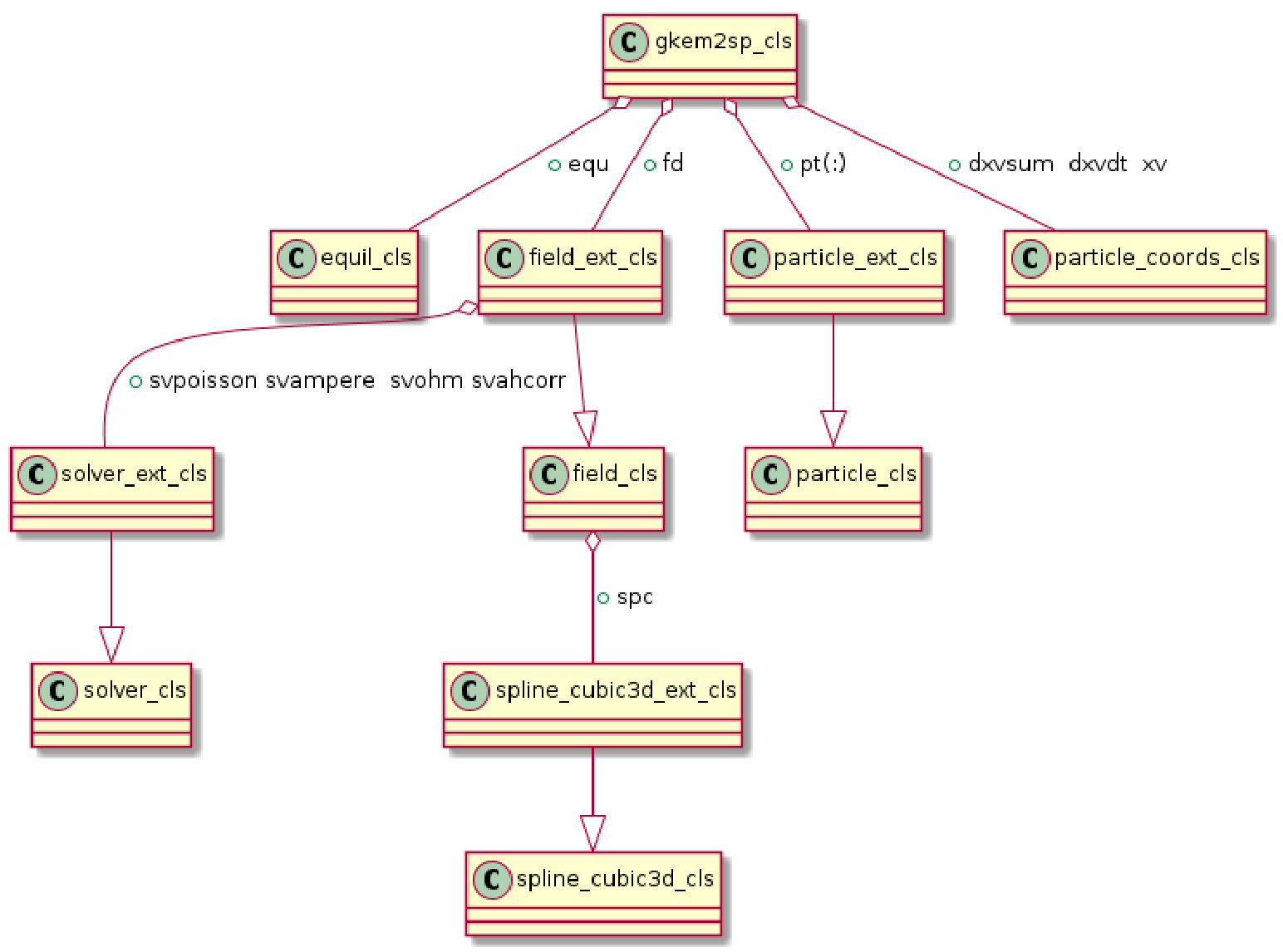}
\caption{The UML (Unified Modeling Language) diagram of GKX code. }
\label{fig:uml}
\end{figure}

\subsection{Normalization}
\label{subsec:normalization}
The normalization of variables in the TRIMEG-GKX code is described in this section. The units used for normalization are denoted with the subscript `N'. The reference length for normalization is $R_{\rm{N}}=1 \; \mathbf{m}$, since the code aims to address not only micro-instabilities but also large-scale modes such as the geodesic acoustic mode. Moreover, using $R_{\rm{N}}=1 \; \mathbf{m}$ allows direct usage of the spatial coordinates from the equilibrium (EQDSK) files without additional normalization. The particle mass is normalized to $m_{\rm{N}}$, with $m_{\rm{N}}=m_{\rm{p}}$ (the proton mass). 
The velocity unit is defined as $v_{\rm{N}}\equiv \sqrt{2T_{\rm{N}}/m_{\rm{N}}}$, 
where $T_{\rm{N}}$ and $m_{\rm{N}}$ are the reference temperature and mass for normalization, respectively.
A typical choice in the TRIMEG code is to use the electron temperature at a specified radial location as $T_{\rm{N}}$ and the proton mass as $m_{\rm {N}}$. 
The time unit is then given by $t_{\rm N}=R_{\rm N}/v_{\rm N}$. The charge unit is the elementary charge $e$. 
Temperatures are normalized by {$T_{\rm{N}}$}.
The magnetic moment $\mu$ is normalized by $v_{\rm{N}}^2/B_{\rm N}$,
where $B_{\rm N}=1\;{\rm T}$ in this work so that the magnetic field data from EQDSK files can be used without additional normalization. 

In addition to the units for the {normalization}, two reference quantities are introduced: the reference density $n_{\rm ref}$ and the reference magnetic field $B_{\rm ref}$. These quantities are used to define two basic parameters: the reference plasma beta $\beta_{\rm ref}$ and the reference Larmor radius ${\rho_{\rm ref} }$, given by
\begin{eqnarray}
    \beta_{\rm ref}=\frac{\mu_0 n_{\rm ref} m_{\rm N} v_{\rm N}^2}{B_{\rm ref}^2} \;\;,  \;\;
    {\rho_{\rm ref} }=\frac{m_{\rm N}v_{\rm N}}{eB_{\rm ref}} \;\;,
\end{eqnarray}
where $\beta_{\rm ref}$ appears in the normalized Amp\`ere's law and ${\rho_{\rm ref} }$ appears in the normalized gyrocenter equation of motion. $\beta_{\rm ref}$ indicates the strength of the electromagnetic effect, and $a/\rho_{\rm ref}$ indicates the normalized device size, where $a$ is the minor radius of the device.
Note that the normalization in TRIMEG differs from that employed in other gyrokinetic codes such as ORB5 \cite{lanti2020orb5} and EUTERPE \cite{kleiber2024euterpe}. In particular, $R_N=1$ m is the length unit since it is the length unit used in EQDSK files which serve as the basis for equilibrium reconstruction in TRIMEG. Meanwhile, $\rho_{\rm{ref}}$ remains an auxiliary variable, as it retains physical significance in expressions related to the magnetic drift velocity and finite Larmor radius effects.

\subsection{The piecewise field-aligned finite element method}
\label{subsec:pfafem}
 The piecewise field-aligned finite element method (PFAFEM) has been developed in our previous work and implemented in the electrostatic particle simulations \cite{lu2025piecewise}. In this work, the PFAFEM has been extended and implemented in the electromagnetic model. It is characterized by two key features: 
\begin{enumerate}
    \item The computational grids are aligned in a traditional pattern without any shift.
    \item The finite element basis functions are defined on piecewise field-aligned coordinates, with each basis function being continuous along the magnetic field line. 
\end{enumerate}
The concept of PFAFEM is summarized as follows \cite{lu2025piecewise}.
We start from the tokamak coordinate $(r,\phi,\theta)$, where $r$, $\phi$ and $\theta$ are the radial-like, toroidal-like and poloidal-like coordinates. The safety factor is $q(r,\theta)\equiv{\bf B}\cdot\nabla\phi/({\bf B}\cdot\nabla\theta)$. The piecewise field-aligned local coordinate $\eta_{k}$ is defined in each toroidal subdomain centered at $\phi_k$ grid as follows,
\begin{eqnarray}
\label{eq:eta_integral}
    \eta_{k}(r,\theta,\phi)= \theta-\int_{\phi_k}^\phi {\rm d}\phi'\frac{1}{q(r,\theta'(r,\phi'),\phi')}  \;\;,
\end{eqnarray}
where the integral is along the magnetic field line and the safety factor $q=q(r,\theta',\phi')$, $\theta'$ is determined by following the magnetic field while varying $\phi'$, namely, $d\theta'/d\phi'=1/q(r,\theta',\phi')$, $\phi_k$ and $\phi$ denote the starting and end points of the integral, respectively. For the straight field line coordinates $r,\bar\theta,\bar\phi$, we have
\begin{eqnarray}
\label{eq:eta_integral_straightB}
    \eta_{k}(r,\bar\theta,\bar\phi)=\bar\theta-\frac{\bar\phi-\bar\phi_k}{\bar q} \;\;,
\end{eqnarray}
where the safety factor $\bar q=\bar q(r)$. The piecewise field-aligned finite element basis function is defined in $(r,\phi,\eta_k)$. The field interpolation at the particle location (``gathering''), the calculation of the perturbed density and parallel current (``scattering'') can be performed in $(r,\phi,\eta_k)$. In addition, the matrices for the field equations are calculated in $(r,\phi,\eta_k)$. PFAFEM is consistent with the theoretical studies of the wave packet in tokamak plasmas \cite{lu2012theoretical}. It can be demonstrated that the partition of unity is satisfied for PFAFEM. More details can be found in the previous work \cite{lu2025piecewise}. 

\subsection{The 3D field-aligned FEM solver and the mixed 2D1F solver}

A mixed particle-in-cell-particle-in-Fourier (PIC-PIF) solver with the finite element method in the radial and poloidal directions but a Fourier representation in the toroidal direction (2D1F) \cite{lu2023full}, and a 3D field-aligned finite element method (FEM) solver are developed in this work to address different types of problems. 
In the 2D1F solver, the FEM is adopted in the $(r,\theta)$ directions, while the Particle-in-Fourier (PIF) scheme is adopted in the toroidal ($\phi$) direction. 
In contrast, the 3D solver applies FEM in all three spatial directions. The size of the grids is $(N_r,N_\theta,N_\phi)$ and $(N_{r,\rm{FEM}},N_{\theta,\rm{FEM}},N_{\phi,\rm{FEM}})$ basis functions are adopted to represent functions in the simulation domain, where $N_{r,\rm{FEM}}=N_r+\Delta N$, $N_{\theta,\rm{FEM}}=N_\theta$, $N_{\phi,\rm{FEM}}=N_\phi$, where $\Delta N=2$ when cubic splines are adopted.  This configuration ensures consistency with the imposed boundary conditions. We apply the periodic boundary conditions in the $(\theta, \phi)$ directions and implement the Dirichlet condition in the $r$ direction. The basis functions are the same as those in our previous work \cite{lu2025piecewise}. 

Four field equations are solved in the GKX code: quasi-neutrality equation, Amp\`ere's law, the iterative Amp\`ere equation, and Ohm's law. The general form of the field equation is
\begin{eqnarray}
\label{eq:field_equation_general}
    M_L\cdot y =   b+M_R\cdot c\;\;,
\end{eqnarray}
where $M_L$ and $M_R$ are partial differential operators, $b$ and $c$ are known vectors, $y$ is the vector to be solved. 
The corresponding general matrix form of the field equations is 
\begin{eqnarray}
\label{eq:mat_poisson}
    \bar{\bar{M}}_{L,ii',jj',kk'} \cdot Y_{i'j'k'}
    =   B^{i,j,k} + \bar{\bar{M}}_{R,ii',jj',kk'} \cdot C_{i'j'k'}\;\;,
\end{eqnarray}
where $Y_{i'j'k'}$ is the field variable to be solved by the linear solver, $B^{i,j,k}$ is from the markers using the projection operator, $C_{i'j'k'}$ is the known field variable, $\bar{\bar{M}}_{L,ii',jj',kk'}$ and $\bar{\bar{M}}_{R,ii',jj',kk'}$ are the matrices on the left- and right-hand sides, respectively. 

For the 3D field-aligned FEM solver, 
\begin{eqnarray}
    \label{eq:field3dfem}
    \bar{\bar{M}}_{L/R,ii',jj',kk'}
    &&= \int {\rm d} r\, {\rm d}\theta\, {\rm d}\phi\, J \Tilde{N}_{ijk} 
    M_{L/R}\Tilde{N}_{i'j'k'}\;\;, \\
    \label{eq:Nijk3dfem}
    B^{i,j,k}
    &&=
    \sum_s C_{{\rm p2g},s}\sum_{p=1}^{N_g} w_p V_p \Tilde{N}_{ijk}(r_p,\eta_{p,k},\phi_p) \;\;, 
\end{eqnarray}
where $\Tilde{N}_{ijk}(r,\eta,\phi)=N_i(r)N_j(\eta_k)N_k(\phi)$, the conversion factor in the projection operator $C_{{\rm p2g},s}=-\Bar{q}_s \langle n\rangle_V V_{\rm tot}/N_g$, $ \langle\ldots\rangle_V$ indicates the volume average, and $V_{\rm tot}$ is the total volume., $V_p$ is a function of the velocity (for the quasi-neutrality equation, $V_p=1$; for Amp\`ere's law, $V_p=v_\|$). 

For the 2D1F solver, we have
\begin{eqnarray}
    \label{eq:field2dfem}
    \bar{\bar{M}}_{L/R,ii',jj',kk'}
    &&=\int {\rm d} r\, {\rm d}\theta\, {\rm d}\phi \, J N_i N_j {\rm e}^{{\text i}k\phi}
    M_{L/R} (N_{i'}N_{j'}{\rm e}^{-{\text i}k\phi})\delta_{k,-k'} \;\;,\\ 
    B^{i,j,k}
    &&=
    \sum_s C_{{\rm p2g},s}\sum_{p=1}^{N_g} w_p V_p N_i(r_p)N_j(\theta_p) {\rm e}^{-{\text i}k\phi_p} \;\;,
\end{eqnarray}
where $N_i=N_i(r)$, $N_j=N_j(\theta)$, $\delta_{i,j}=1$ if $i=j$, $\delta_{i,j}=0$ if $i\ne j$, and $V_p$ is the same as that in Eq.\eqref{eq:Nijk3dfem}.
The grid variables $Y_{ijk}$ and $B^{ijk}$ interact with the particles in the ``scatter'' and ``gather'' processes. 
The scatter operation assigns the charge and current densities back to the grid. 
In the gather operation, field values are interpolated at the position of the particles.
In the scattering process using the 3D field-aligned FEM, the values of the basis function $\tilde{N}_{ijk}(r,\eta,\phi)$ is calculated at the particle location, but in $(r,\eta,\phi)$ coordinates as shown in Eq.~\eqref{eq:Nijk3dfem}. Thus, the grid value $Y^{ijk}$ is obtained by taking into account all particles where $\tilde{N}_{ijk}(r,\eta,\phi)$ does not vanish. 
In the gathering process using the 3D field-aligned FEM, the field at the particle location is
\begin{eqnarray}
Y(r_p,\theta_p,\phi_p)=\sum_{i,j,k} Y_{i,j,k} N_i(r_p)N_j(\eta_{k,p})N_k(\phi_p)\;\;. 
\end{eqnarray}
The B-spline coefficients of $\eta(r,\theta,\phi)$ are stored in a three-dimensional matrix in the TRIMEG code as it also applies to the open field line region when the $R,\phi,Z$ coordinates are used and the three-dimensional interpolation is evoked in both the ``gather'' and ``scatter'' processes. As a future optimization for core plasma simulations using a numerical equilibrium, $\eta$ can be reduced to $\eta=\bar\theta-\phi/\bar{q}$ in the straight field line coordinate $(r,\phi,\bar\theta)$ where $\bar{q}\equiv{\bf B}\cdot\nabla\phi/{\bf B}\cdot\nabla\bar\theta$ is independent of $\bar\theta$, which is more efficient as only a one-dimensional interpolation of $\bar{q}(r_p)$ is needed. 

\subsection{Interface to the Experimental Data Structure}
GKX interfaces directly with the EQDSK equilibrium files, which define the poloidal flux as a function of 
$(R,Z)$. The magnetic equilibrium field B is internally reconstructed using spline interpolation in GKX. For the experimental EQDSK data, GKX employs a radial coordinate defined as the square root of the normalized poloidal flux, $r=\sqrt{(\psi-\psi_{\rm axis})/(\psi_{\rm edge}-\psi_{\rm axis})}$. For the \textit{ad-hoc} equilibrium with concentric circular cross-section, GKX employs an analytical $q$ profile, and $\rho=\sqrt{(R-R_0)^2+(Z-Z_0)^2}$ is adopted as the radial coordinate since it is a more efficient way. Density and temperature profiles can be supplied directly in this coordinate system. If provided in alternative coordinates (e.g., real-space or toroidal magnetic flux-based), the code can transform them to the required form using the equilibrium data. 

GKX provides an interface to the IMAS (Integrated Modelling \& Analysis Suite) data framework, developed for ITER, which adopts a toroidal magnetic flux coordinate system. IMAS comprises a comprehensive suite of infrastructure components, physics modules, and analysis tools. Within IMAS, data is organized into Data Entries, where each Data Entry represents a collection of Interface Data Structures (IDSs) grouped to form a unified dataset. 

To facilitate simulations, GKX includes scripts for extracting data from various IDSs and converting it into the standard input format required by the code. For example, the JET multi-species gyrokinetic case in Section \ref{subsec:jet}  is initialized using this data processing workflow. The \texttt{equilibrium} IDS describes magnetic equilibrium information following the COCOS = 17 coordinate convention \cite{sauter2013cocos}, and can be converted into an EQDSK file format using the CHEASE code \cite{lutjens1996chease}.

The density and temperature data at different time slices are stored in the \texttt{core\_profiles} IDS as one-dimensional arrays, defined as functions of the toroidal magnetic flux coordinate. To express these profiles in terms of the poloidal flux, a mapping between the toroidal and poloidal magnetic flux surfaces is computed. Using this mapping, the profiles are interpolated using cubic interpolation and reformulated as functions of the normalized flux coordinate, defined by \( r = \sqrt{(\psi - \psi_{\mathrm{axis}})/(\psi_{\mathrm{edge}} - \psi_{\mathrm{axis}})} \). More generally, the profiles of the parallel velocity and the radial electric field can also be extracted and processed in the same manner. These quantities will be incorporated in future developments of GKX to support the inclusion of neoclassical physics effects in the simulation framework. 

\subsection[Strong]{Strong form of $\partial_t\delta{A}_\|^{\rm s}+\partial_\|\delta\Phi$ for rigorous treatment of GC motion}
\label{subsec:strong_AP}
Noise reduction is critical in the PIC simulations. According to the ideal Ohm's law Eq.~(\ref{eq:ohm_law0}), the term $\partial_\| \delta\Phi + \partial_t\delta A_\|^{\rm s}$ should vanish in the gyrocenter equation of motion $\delta \dot u_\|$ theoretically. However, numerically, the coefficients of the basis functions are solved for $\delta\Phi$ and $\partial_t\delta{A}^{\rm s}$ in Eq.~(\ref{eq:mat_poisson}) and only the weak form of $\partial_\| \delta\Phi + \partial_t\delta A_\|^{\rm s}=0$ is satisfied. At an arbitrary location, $\partial_\| \delta\Phi + \partial_t\delta A_\|^{\rm s}$ is finite (not exactly zero), due to the approximation embedded in the finite element representation. In pushing the gyrocenters, errors can accumulate in $\delta \dot u_\|=(e_s/m_s)\langle\partial_\| \delta\Phi + \partial_t\delta A_\|^{\rm s}\rangle+\ldots=\sigma+\dots$ without corrections, where $\sigma$ indicates the finite numerical error at the particle location. To improve the accuracy, GKX does not explicitly enforce the condition $\partial_\| \delta\Phi + \partial_t \delta A_\|^{\rm s} = 0$. Instead, both $\partial_\| \delta\Phi$ and $\partial_t\delta A_\|^{\rm s}$ are retained separately in the formulation. Consequently, we use the numerical representation of $\partial_\|\delta\Phi$ and $\partial_t\delta A^{\rm s}_\|$ in the gyrocenter equations of motion with the following form without relying on their analytical cancellation,
\begin{eqnarray}
	\delta \Phi(r,\phi,\theta) &=& \sum_{i,j,k}\delta\Phi_{i,j,k} N_i(r) N_j(\theta) N_k(\phi)   \;\;, \\
	\partial_t\delta A^{\rm s}_\|(R,\phi,Z) &=& \sum_{i,j,k}(\partial_t\delta A^{\rm s}_{\|})_{i,j,k} N_i(r) N_j(\theta) N_k(\phi)   \;\;.
\end{eqnarray}
This is referred to as using the strong form, where both $\partial_\|\delta\Phi$ and $\partial_t \delta A_\|^{\rm s}$ are numerically represented and evaluated explicitly. The coefficient $\delta\Phi_{i,j,k}$ is from the weak form of the quasi-neutrality equation while $(\partial_t\delta A^{\rm s}_{\|})_{i,j,k}$ is directly from the weak form of the Ohm's law. This leads to a more robust and accurate simulation, especially when the noise-mitigation tricks like Fourier filtering are not applied. Since we avoid using numerical buffers near the simulation boundary or the Fourier filters, the rigorous treatment can improve the accuracy and the signal-to-noise ratio even without filters or buffers. 

\subsection{Cache optimization for particle pusher}
Cache optimization can significantly impact performance. Different methods can improve the cache access, such as particle sorting and the optimization of particle/field data layouts. In the GKX code,  the most computationally intensive part is the particle pusher. Particles are advanced one by one in the particle loop, and both the equilibrium and perturbed fields are calculated within the loop before pushing the particle forward. The procedure is described by the following pseudocode.

{\footnotesize
\begin{verbatim}
do i=1,nptot
    pt_radius= particle%radius(i)
    pt_theta = particle%theta(i)
    pt_phi = particle%phi(i)
    pt_Apar = field%calc_field(field%apar,pt_radius,pt_theta,pt_phi)
    pt_Phi  = field%calc_field(field%phi,pt_radius,pt_theta,pt_phi)
    pt_B   = equ%calc_B(pt_radius,pt_theta)
    particle%onestep(pt_radius,pt_theta,pt_phi, pt_Apar, pt_Phi, pt_B)
enddo
\end{verbatim}
}

It is worth noting that in the GKX code, additional field derivatives need to be calculated, such as $\partial_\alpha\delta\Phi$, $\partial_\alpha\delta A^{\rm s}_\|$ and $\partial_\alpha\delta A^{\rm h}_\|$, where $\alpha\in(r,\theta,\phi)$. The optimization is achieved by moving the calculation of the perturbed field outside of the particle loop.

{\footnotesize
\begin{verbatim}
do i=1,nptot
    pt_Apar1d(i) = field%calc_field( &
      field%apar,particle%radius(i),particle%theta(i),particle%phi(i))
enddo
do i=1,nptot
    pt_Phi1d(i)  = field%calc_field( &
      field%phi,particle%radius(i),particle%theta(i),particle%phi(i))
enddo
do i=1,nptot
    pt_radius= particle%radius(i)
    pt_theta = particle%theta(i)
    pt_phi = particle%phi(i)
    pt_Apar = pt_Apar(i)
    pt_Phi  = pt_Phi(i)
    pt_B   = equ%calc_B(pt_radius,pt_theta)
    particle%onestep(pt_radius,pt_theta,pt_phi, pt_Apar, pt_Phi, pt_B)
enddo
\end{verbatim}
}

The effects of the cache optimization on speed-up depend on the parameters of the simulations, such as the type of field solver, the marker number, and the grid size. A typical single harmonic simulation adopted in Section \ref{subsec:kbm_gknet} is featured by $16\times10^6$ electrons, $10^6$ ions (with 4-point gyro average),  and grid resolutions of order $N_r=96$, $N_\theta=128$. It takes  292.5 and 3921.3 seconds to run 200 steps with and without the cache optimization, respectively. The overall speed-up is $13.4$. The details of the speed-up using this optimization are shown in Fig.~\ref{fig:pt_fd_comparison}. The cache optimization speeds up not only the particle pusher and the scattering (g2p) operations, but also the recording and pullback suboutines, where the field is also calculated.

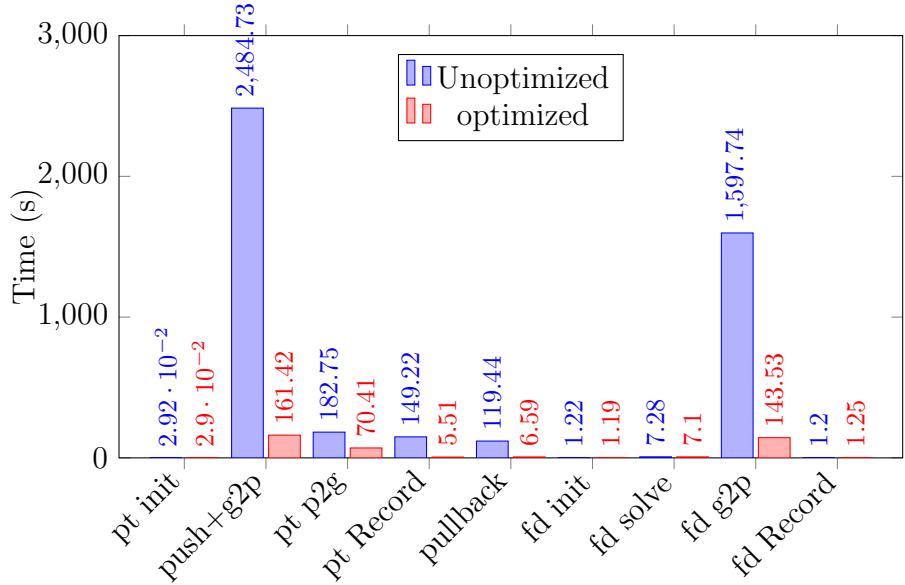
\begin{figure}[htbp]
\centering
\begin{tikzpicture}
\begin{axis}[
    ybar,
    bar width=12pt,
    ylabel={Time (s)},
    width=12cm,
    height=7.2cm,
    symbolic x coords={
        pt init, push+g2p, pt p2g, pt Record, pullback,
        fd init, fd solve, fd g2p, fd Record
    },
    xtick=data,
    x tick label style={rotate=45, anchor=east},
    ymin=0,
    ymax=3000, 
    enlarge x limits=0.1,
    nodes near coords,
    every node near coord/.append style={font=\footnotesize , rotate=90, anchor=west},
    legend style={at={(0.5,0.75)}, anchor=south, legend columns=2,legend columns=1}
]
\addplot coordinates {
    (pt init, 0.0292)
    (push+g2p, 2484.726 )
    (pt p2g,  182.749)
    (pt Record, 149.221342)
    (pullback, 119.435)
    (fd init, 1.222)
    (fd solve, 7.276)
    (fd g2p, 1597.741)
    (fd Record, 1.201)
};
\addplot coordinates {
    (pt init, 0.029)
    (push+g2p, 161.418)
    (pt p2g,70.409)
    (pt Record,5.512 )
    (pullback, 6.593)
    (fd init, 1.186)
    (fd solve, 7.097)
    (fd g2p, 143.53)
    (fd Record, 1.248)
};
\legend{Unoptimized, optimized}
\end{axis}
\end{tikzpicture}
\caption{Comparison of execution times for Particle subroutines (PT) and Field subroutines (FD). PT times are summed over two species.  It takes  292.5 and 3921.3 seconds to run 200 steps with and without the cache optimization, respectively. ``g2p'' denotes the field interpolation (gathering) and ``g2p'' denotes the scattering operation.  Note that ``g2p'' is intensively used in ``push'' and thus ``push+g2p'' appears as an item. }
\label{fig:pt_fd_comparison}
\end{figure}

\section{Simulation results}
\label{sec:results}
In this section, we demonstrate the efficiency of our code in solving electromagnetic problems. For the first time, object-oriented programming, the Piecewise Field-Aligned Finite Element Method (PFAFEM), cache optimization, and a rigorous treatment of the strong form in $\delta\dot{u_\|}$ have been integrated into a single code framework to achieve high performance. Previous physics applications of the TRIMEG-GKX/C0 code can be found in~\cite{guo2022mode,lana2023neoclassical}. The convergence of the iterative Ampère solver, electrostatic multi-$n$ nonlinear simulations, and studies employing Fourier filters were demonstrated in our earlier works~\cite{lu2023full,lu2025piecewise}.

Here, we present results related to recent developments, including simulations of energetic particle-driven Alfvén eigenmodes, ion temperature gradient (ITG) modes, and kinetic ballooning modes (KBMs), performed on both ad hoc and experimental equilibria. Specifically, Section~\ref{subsec:jet} demonstrates the excellent computational scalability of multi-species treatment in simulations of plasma in the Joint European Torus (JET).
Section~\ref{subsec:itpatae} highlights the code's capability for electromagnetic simulations in the small electron skin depth limit within an ad hoc equilibrium.
Section~\ref{subsec:nled} presents simulations of energetic particle-driven reversed-shear-induced Alfvén eigenmodes in ASDEX-Upgrade plasmas.
Section~\ref{subsec:cbc_itgkbm} shows ITG and KBM simulations in the Cyclone Base Case, with results in good agreement with the GENE code.
Section~\ref{subsec:tcv} demonstrates ITG mode simulations in the Tokamak à Configuration Variable (TCV).
Section~\ref{subsec:kbm_gknet} presents single- and multiple-harmonic simulations of ITG modes and KBMs, demonstrating the code’s performance in high-$\beta$ plasmas.
In addition to extensive linear benchmarks, several nonlinear simulation results are also presented in Sections~\ref{subsec:tcv} and~\ref{subsec:kbm_gknet}.


\subsection{Computational performance of multi-species simulations (JET plasma)}
\label{subsec:jet}
Since object-oriented programming has been integrated into the GKX code as shown in Fig.~\ref{fig:uml}, multiple species can be readily treated by initializing instances for all the species. One key issue is the computational consumption as the number of species increases. The JET (Joint European Torus) case is adopted for this study since multiple species are relevant for the identification of the physics. Specifically, the stable Deuterium-Tritium plasmas of discharge \#99896 with improved confinement have been reported in the presence of energetic-ion instabilities \cite{garcia2024stable}. Nine species are considered for their possible effects on the turbulence evolution and the interaction with the Alfv\'en modes. These species include the electron, three thermal ions (Hydrogen, Deuterium, Tritium), three fast ions (Hydrogen, Deuterium, Helium), and two impurity species (Beryllium and Nickel). 

The gyrokinetic simulations are performed in the GKX code using 3, 4, 5, 6, 7 and 9 species for the $n=5$ Toroidicity induced Alfv\'en Eigenmode (TAE). A Maxwellian distribution is assumed for all species. We use $16\times10^6$ electron markers and $10^6$ ion markers per species without gyro-average. A typical 2D mode structure is shown in the left frame of Fig.~\ref{fig:jet_multispecies}. To study the scalability for the specific number of species, a set of simulations are run for $\sim10$ hours on 2 nodes (AMD EPYC Genoa 9554) of the MPCDF Viper supercomputer, with 128 CPU cores on each node, with a processor base frequency of $3.1$ GHz and a max turbo frequency of $3.75$ GHz. The computational efficiency is demonstrated in the right frame of Fig.~\ref{fig:jet_multispecies} in terms of seconds per step. The total particle number for 3, 4, 5, 6, 7, and 9 species is $N_p=(18, 19, 20, 21, 22, 24)\times10^6$ , respectively, and thus the normalized efficiency is calculated by multiplying with the factor $N_p/18\times10^6$. The results show that normalized efficiency remains nearly constant as additional species are included, demonstrating the good scalability with respect to the species count. The physics studies will be reported in a separate work in the future. 

\begin{figure}[h!]
\centering
\begin{minipage}{0.4\textwidth}
    \centering
    \includegraphics[width=\textwidth]{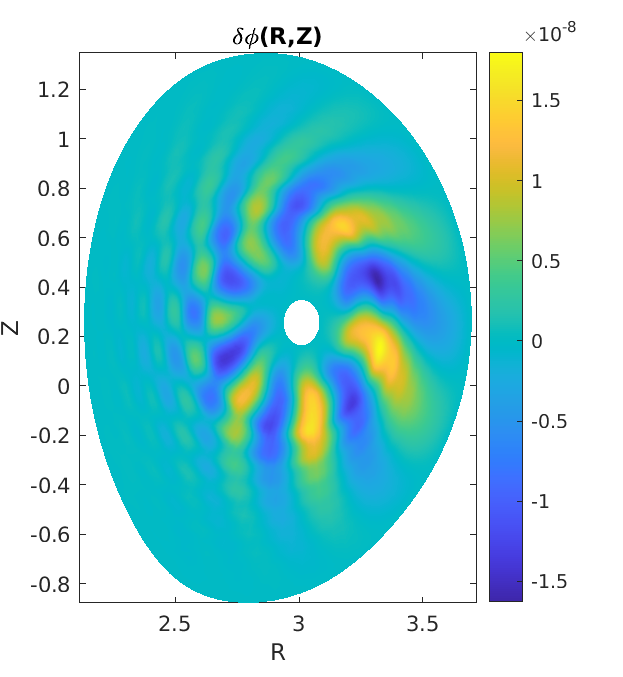}
\end{minipage}
\hfill
\begin{minipage}{0.58\textwidth}
    \centering
    \begin{tikzpicture}
    \begin{axis}[
        width=\textwidth,
        height=6cm,
        xlabel={Species},
        ylabel={Time (sec/step or normalized)},
        symbolic x coords={3,4,5,6,7,9},
        xtick=data,
        ymin=1.5,
        legend style={
            at={(0.97,0.03)},
            anchor=south east,
            draw=none,
            fill=white,
            font=\small
        },
        grid=major,
        enlarge x limits=0.1,
    ]
    \addplot+[
        mark=*,
        nodes near coords,
        every node near coord/.append style={yshift=0pt}
    ] coordinates {
        (3,2.57) (4,2.56) (5,2.72) (6,2.96) (7,3.01) (9,3.39)
    };
    \addlegendentry{sec/step}

    \addplot+[
        mark=square*,
        color=red,
        nodes near coords,
        every node near coord/.append style={yshift=-15pt}
    ] coordinates {
        (3,2.57) (4,2.4253) (5,2.4480) (6,2.5371) (7,2.4627) (9,2.5425)
    };
    \addlegendentry{normalized}
    \end{axis}
    \end{tikzpicture}
\end{minipage}
\caption{\label{fig:jet_multispecies}The 2D mode structure of $\delta\Phi$ from the JET simulation plasma in Section \ref{subsec:jet} (left). Computation time per step and normalized values for different species counts (right).}
\end{figure}
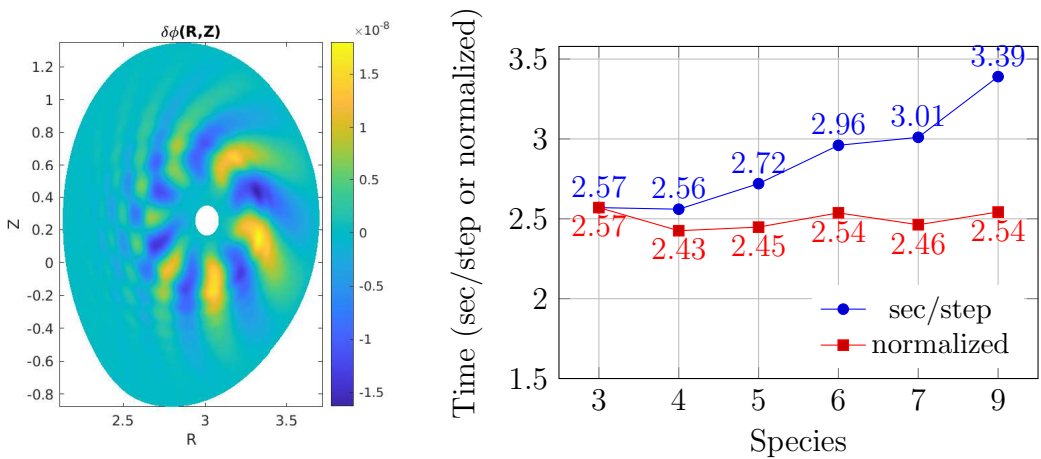


\subsection{Energetic particle driven Toroidicity induced Alfv\'en Eigenmodes (ITPA-TAE)}
\label{subsec:itpatae}
The toroidicity induced Alfv\'en eigenmode driven by energetic particles is simulated using the parameters defined by the ITPA-EP (International Tokamak Physics Activity-Energetic Particle) group \cite{konies2018benchmark}. 
For this case, the major radius $R_0=10 \; {\rm m}$, minor radius $a=1$ m, on-axis magnetic field $B_{\rm axis}=3$ T, and the safety factor profile $q(r)=1.71+0.16r^2$. The electron density and temperature are constant with $n_{{\rm e}0}=2.0\times10^{19}\;\rm{m}^{-3}$, $T_{\rm{e}}=1$~keV. The ratio of the electron pressure to the magnetic pressure is $\beta_{\rm e}\approx 9\times 10^{-4}$. The Larmor radius of the thermal {ions} is $\rho_{\rm{ti}}=m_{\rm{i}}v_{\rm{ti}}/(eB_{\rm{axis}})=1.52\times10^{-3}$~m. The nominal mass ratio $m_{\rm i}/m_{\rm e}=1836$ is used, and the ratio between the adiabatic part ($\delta A_\|^{\rm h}/d_{\rm e}^2$) and the non-adiabatic part ($\nabla_\perp^2\delta A^{\rm{h}}_\|$) in the left-hand side of Amp\`ere's equation is $1/(d_{\rm e}^2k_\perp^2)\approx \beta_{\rm e}/(k_\perp\rho_{\rm{ti}})^2(m_{\rm{i}}T_{\rm i}/m_{\rm{e}}T_{\rm{e}})\approx1.622\times10^3$, where $k_\perp\approx nq/r=6\times1.75/0.5=21$.  This ITPA-TAE case is featured with a small electron skin depth ($d_{\rm e}\approx1.182\times10^{-3}$ m) and suffers from the ``cancellation problem'' if the pullback scheme is not adopted. Note that the poloidal Fourier filter has not been used in this work.  Consequently, the time-step size ($\Delta t$) is smaller (the maximum $\Delta t\propto k_\| v_{\rm A} $, where $k_\|$ is the parallel wave vector in the system, $ v_{\rm A} $ is the Alfv\'en velocity) than that with structured meshes and the Fourier filter \cite{lu2023full} by a factor of $\sim1/50$.
 
 The EP density profile is given by 
\begin{eqnarray}
\label{eq:nEP1d}
	n_{\rm{EP}}(r)&=&n_{\rm{EP},0}c_3 \exp\left[ -\frac{c_2}{c_1} \tanh\left(\frac{{r}-c_0}{c_2}\right)\right]\;\;, \\
	\frac{{\rm d}\ln n_{\rm{EP}}}{{\rm d} r} &=&\cosh^{-2}\left(\frac{ r-c_0}{c_2}\right)\;\;,
\end{eqnarray}
where the normalized radial-like coordinate $r=\sqrt{(\psi-\psi_{\rm axis})/(\psi_{\rm edge}-\psi_{\rm axis})}$, $n_{\rm{EP},0} =1.44131\times10^{17}\; \rm{m}^{-3}$, the subscript `$\rm{EP}$' indicates EPs (energetic particles), $c_0 = 0.491 23$, $c_1 =0.298 228$, $c_2 =0.198 739$, $c_3 =0.521 298$. For the base case, the EP temperature is $400\; {\rm keV}$. The $n=6$ mode is simulated by applying a toroidal Fourier filter. The initial perturbation in the marker weights is applied with two poloidal harmonics with $m=10$ and $11$. A total of 
 $32\times10^6$ electron markers, $4\times10^6$ ion markers and $4\times10^6$ energetic particle markers are simulated. The radial grid number is $N_r=64$.  

The radial structure of the poloidal harmonics for $T_{\rm EP}=400\; {\rm keV}$ is shown in the left frame of Fig.~\ref{fig:mode2d_itpa}. The agreement with the EUTERPE results is reasonably good (the results from ORB5, GYGLES, and other codes can be found in \cite{konies2018benchmark}). 
The growth rate for different values of $T_{\rm EP}$ is shown in the right frame. The GKX result follows the trend of those from ORB5 and GYGLES. It should be noted that there are different treatments/simplifications for the higher-order terms in the gyrocenter equations of motion, the lost particles, and the gyro-average of the gradient among different codes, which can lead to differences by several percent of the growth rate. Nevertheless, it is indicated that GKX agrees reasonably with the other two codes in  simulations of EP-driven TAEs. 

\begin{figure}[t]
\centering
\includegraphics[width=0.45\textwidth]{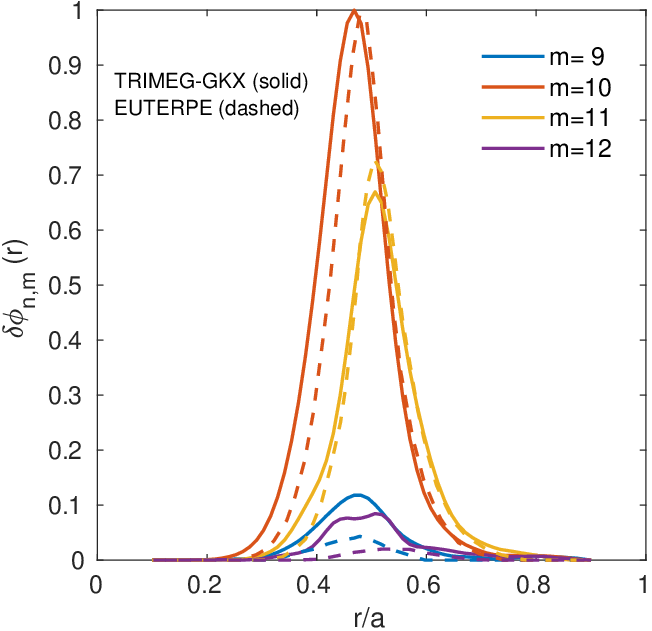}
\includegraphics[width=0.47\textwidth]{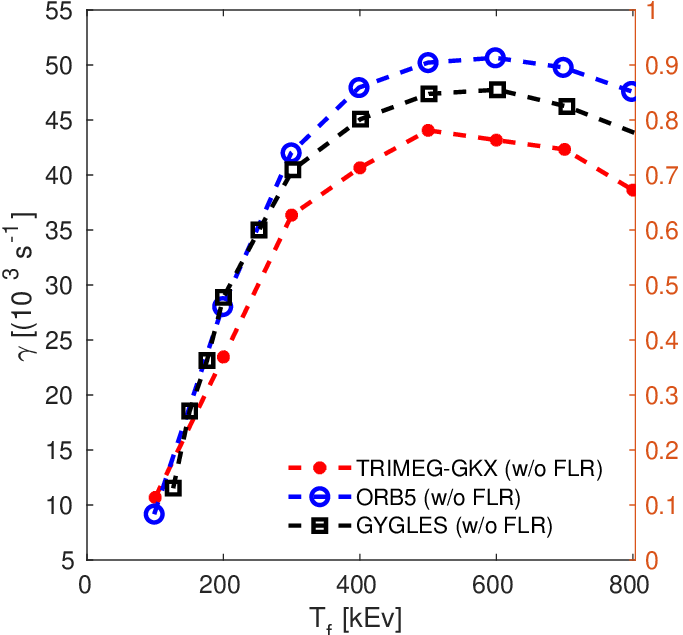}
\caption{The energetic particle driven Toroidicity induced Alfv\'en Eigenmode (TAE) in ad-hoc equilibrium in Section \ref{subsec:itpatae}. The radial structure of the poloidal harmonics (left). The growth rate versus the energetic particle temperature (right).}
\label{fig:mode2d_itpa}
\end{figure}

\subsection{EP driven Reversed Shear Alfv\'en Eigenmodes (RSAE) in ASDEX-Upgrade (NLED-AUG)}
\label{subsec:nled}
In this section, we use the realistic geometry of the ASDEX Upgrade (AUG) tokamak with discharge number \#31213 at t= 0.84~s. This is a typical discharge for the studies of energetic particle physics in different codes and models  \cite{lauber2018strongly}. The benchmark has been reported recently \cite{vlad2021linear}. The density and temperature profiles of the electrons, thermal ions, and fast particles are shown in Fig.~\ref{fig:nTq_nled}. 

We  consider only finite orbit width (FOW) effects when describing EPs, neglecting finite Larmor radius (FLR) ones to compare with the results from other codes \cite{vlad2021linear}. The electron, thermal ion, and fast ion marker numbers are $16\times10^6, 10^6, 10^6$, respectively. The ion-to-electron mass ratio used in this case is $m_{\rm i}/m_{\rm e}=500$.  The time-step size is $\Delta t/t_N=0.004$, where $t_N$ is the code time unit defined in Section \ref{subsec:normalization}.
The mode structure in a 2D cross-section and the radial structure of the poloidal Fourier harmonics are shown in Fig.~\ref{fig:mode2d_nled}. The 2D mode structure is similar to those simulated using the HYMAGYC code, and the radial structure of the poloidal harmonics is consistent with those from the ORB5 codes reported previously \cite{vlad2021linear}. 
Fig.~\ref{fig:growth_nled} shows the growth rate versus the EP temperature. The results from ORB5 and GKX are in better agreement with each other compared to the other codes, since ORB5 and GKX both adopt a gyrokinetic plasma model, while the MEGA and HYMAGYC are hybrid MHD-kinetic codes. 

\begin{figure}[t]
\centering
\includegraphics[width=0.98\textwidth]{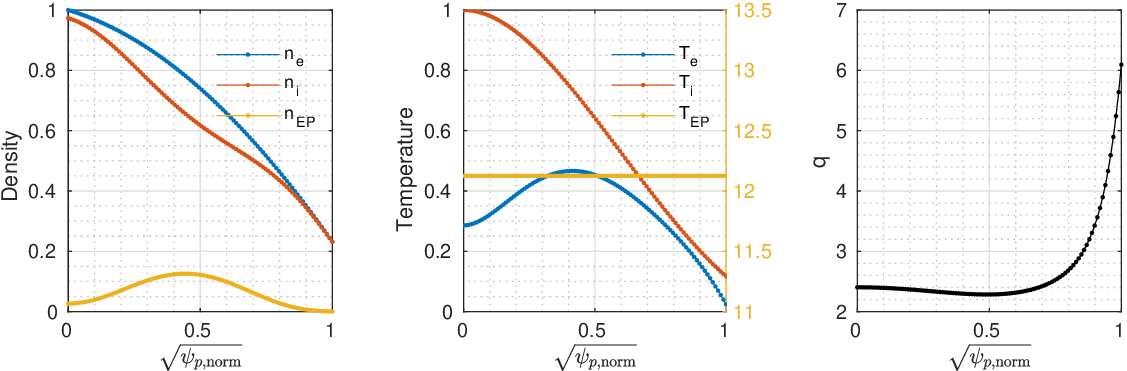}
\caption{The density (left), temperature (middle) profiles of the electrons, thermal ions, and fast particles of the NLED-AUG case in Section \ref{subsec:nled}. The density and temperature are normalized to $n_{\rm ref}=1.7159\times10^{19}/{\rm m^3}$ and $T_{\rm ref}=2.4741\times10^3$~eV, respectively. The radial profile of the safety factor (right). }
\label{fig:nTq_nled}
\end{figure}

\begin{figure}[t]
\centering
\includegraphics[width=0.46\textwidth]{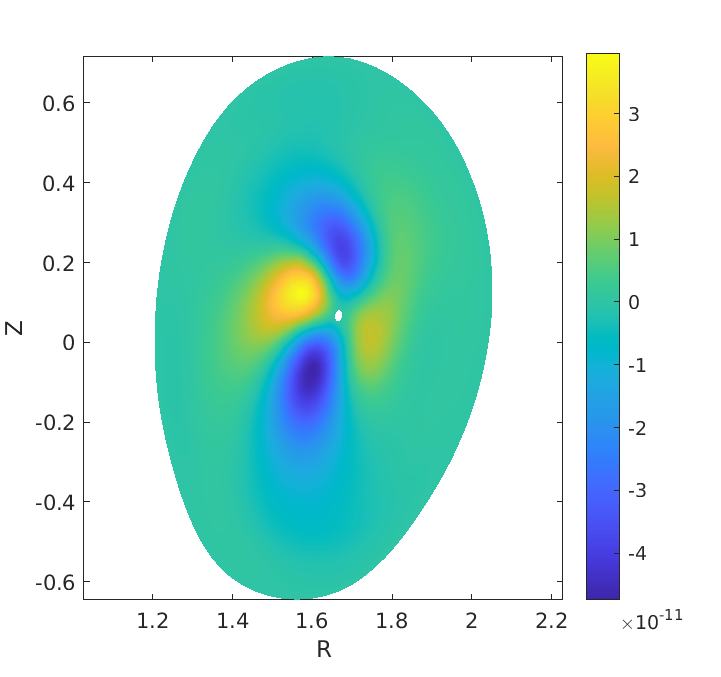}
\includegraphics[width=0.42\textwidth]{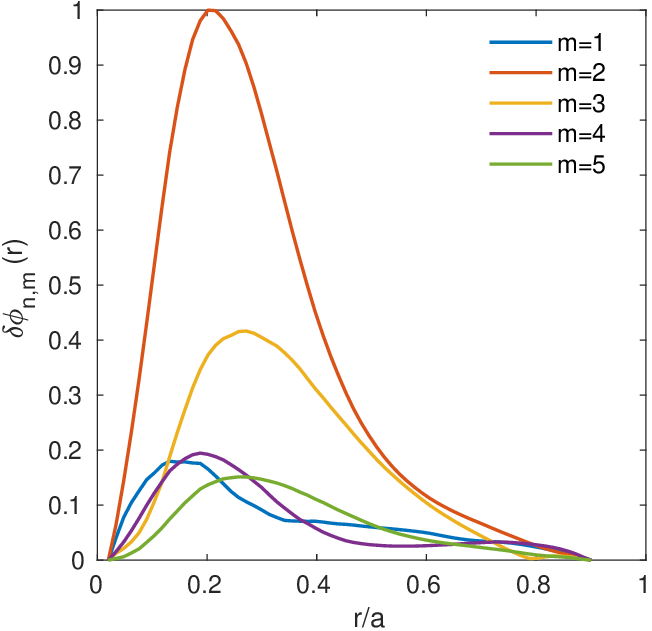}
\caption{The 2D mode structure (left) and the radial structure of the poloidal harmonics (right) of the Reversed Shear Alfv\'en Eigenmode (RSAE) in ASDEX-Upgrade tokamak plasma studied in Section \ref{subsec:nled}.}
\label{fig:mode2d_nled}
\end{figure}

\begin{figure}[t]
\centering
\includegraphics[width=0.65\textwidth]{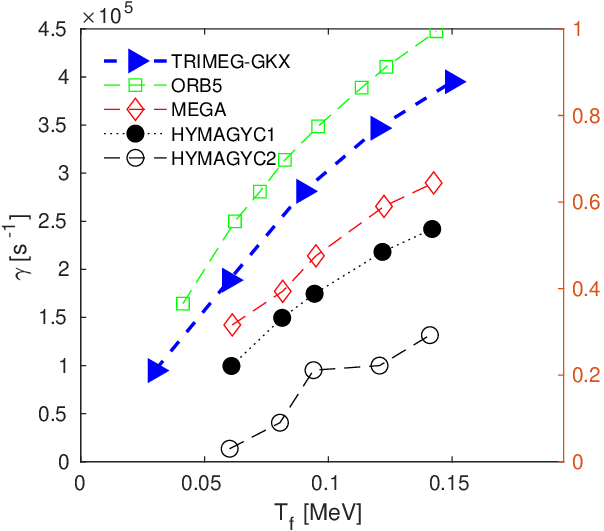}
\caption{Growth rate versus energetic particle temperature obtained from GKX and other codes \cite{vlad2021linear} studied in Section \ref{subsec:nled}. }
\label{fig:growth_nled}
\end{figure}

\subsection{Ion temperature gradient mode and kinetic ballooning mode (Cyclone)}
\label{subsec:cbc_itgkbm}
The GA-STD case (core) parameters are based on the DIII-D tokamak plasma and are adopted in the previous benchmark work \cite{gorler2016intercode}. Concentric circular magnetic flux surfaces are adopted. The equilibrium density and temperature profiles denoted as $H(\tilde r)$, and the normalized logarithmic gradients $R_0/L_\mathrm{H}$, are given by 
\begin{eqnarray}
\label{eq:nTprofile}
     \frac{H(\tilde r)}{H(\tilde r_0)}  =\exp\left[-\kappa_\mathrm{H}w_\mathrm{H}\frac{a}{R_0}\tanh\left(\frac{\tilde r-\tilde r_0}{w_\mathrm{H}a}\right)\right]\;\;,  
\end{eqnarray}
where $\tilde r=\sqrt{(R-R_{\rm axis})^2+(Z-Z_{\rm axis})^2}$, $L_\mathrm{H}=-\left[{\rm d}\ln H(\tilde r)/{\rm d}\tilde r\right]^{-1}$ is the characteristic length of profile $H(\tilde r)$ and $\tilde r_0=a/2$. 
The ion-to-electron mass ratio is $m_{\rm i}/m_{\rm e}=1836$ and the deuterium ($m_{\rm i}/m_{\rm p}=2$) is the only ion species where $m_{\rm p}$ is the mass of a proton.
The on-axis magnetic field $B_0=2\rm{T}$, $\rho^*\equiv\rho_{\rm i}/a=1/180$, $\rho_{\rm i}=\sqrt{2T_{\rm i}m_{\rm i}}/(eB_0)$, aspect ratio $\epsilon=a/R_0=0.36$, $T_{\rm e}/T_{\rm i}=1$, characteristic length of temperature and density profiles $R_0/L_{T_{\rm i}}=-(\rm{d}\,\rm{ln }T_{\rm i}/\rm{d}\tilde r)^{-1}=6.96$, $R_0/L_{T_{\rm e}}$$=-(\rm{d}\ln T_{\rm e}/\rm{d}\tilde r)^{-1}=6.96$, $R_0/L_{n}=-({\rm d}\ln n/{\rm d}\tilde r)^{-1}=2.23$, and collision frequency $\nu_{\rm coll}=0$.

The simulation domain is $r/a\in[0.1,0.9]$. For the low $\beta$ case, $256\times10^6$ electrons and $4\times10^6$ ions are simulated. Since the pullback-mixed variable scheme performs better at the MHD limit, fewer electrons are needed for higher $\beta$ cases. For the high $\beta$ cases, $128\times10^6$ electrons and $10^6$ ions are used. The time-step size is $0.0005 t_N$.  The simulation is run on two (for $128\times10^6$ electrons) or four (for $256\times10^6$ electrons) nodes of the supercomputer Viper. A typical simulation runs for 24$\sim$48 hours to cover at least $10 R_N/v_N$ until the exponential growth is observed for the calculation of the growth rate. 

The growth rate at different values of $\beta$ is shown in Tab. \ref{tab:growth_cbc}. A good agreement is achieved between the GKX result and the GENE result \cite{gorler2016intercode}. 
Figure \ref{fig:mode2d_cbc} shows the 2D mode structures of the scalar potential $\delta\Phi$ and the parallel vector potential $\delta{A}_\|$ for the simulation with a single toroidal harmonic $n=19$. The ITG mode structure at $\beta=0.001$ is similar to that at $\beta=0.025$, in terms of the ballooning structure in $\delta\Phi$. However, the ratio between the maximum magnitude of $\delta\Phi$ and $\delta{A}_\|$ is different. For ITG at $\beta=0.001$, $|\delta\Phi/\delta{A}_\| |\gtrsim 50$. For KBM at $\beta=0.025$, $|\delta\Phi/\delta{A}_\|  |\approx 4$. This trend is consistent with the previous ITG-KBM simulations in GKNET \cite{ishizawa2019global}. 


\begin{table}[h!]
\centering
\begin{tabular}{c c c}
$\beta$ & $\gamma [c_s/R_0]$ (GKX) &  $\gamma [c_s/R_0]$ (GENE) \\
\hline 
0.001 & 0.533 & 0.528 \\
0.020 & 1.012 & 0.989 \\
0.025 & 1.442 & 1.467 \\
\end{tabular}
\caption{\label{tab:growth_cbc}Comparison of growth rates for GKX and GENE \cite{gorler2016intercode} at different values of $\beta$.}
\end{table}

\begin{figure}[t]
\centering
\includegraphics[width=0.4\textwidth]{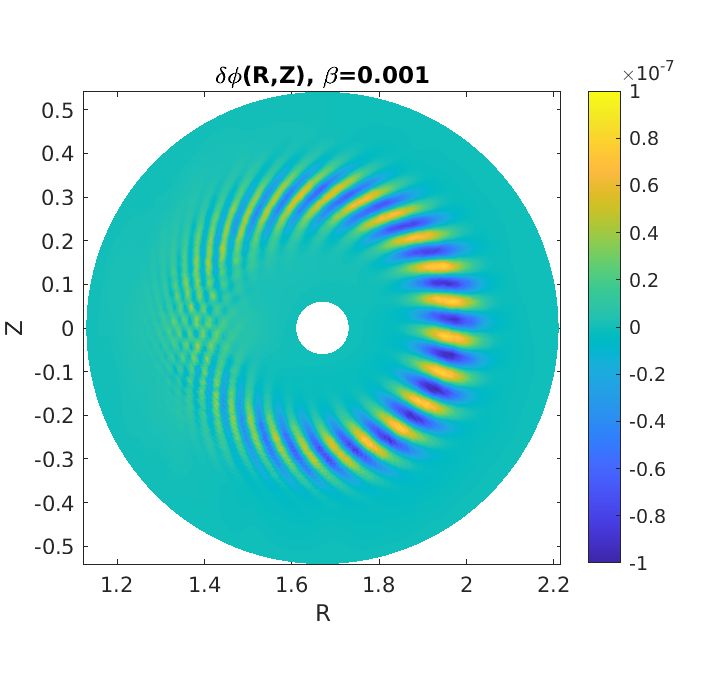}
\includegraphics[width=0.4\textwidth]{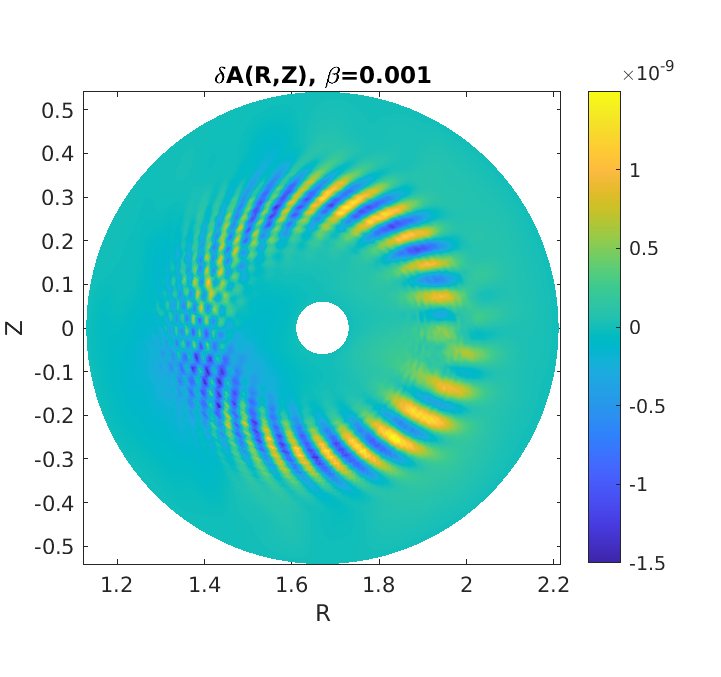}
\includegraphics[width=0.4\textwidth]{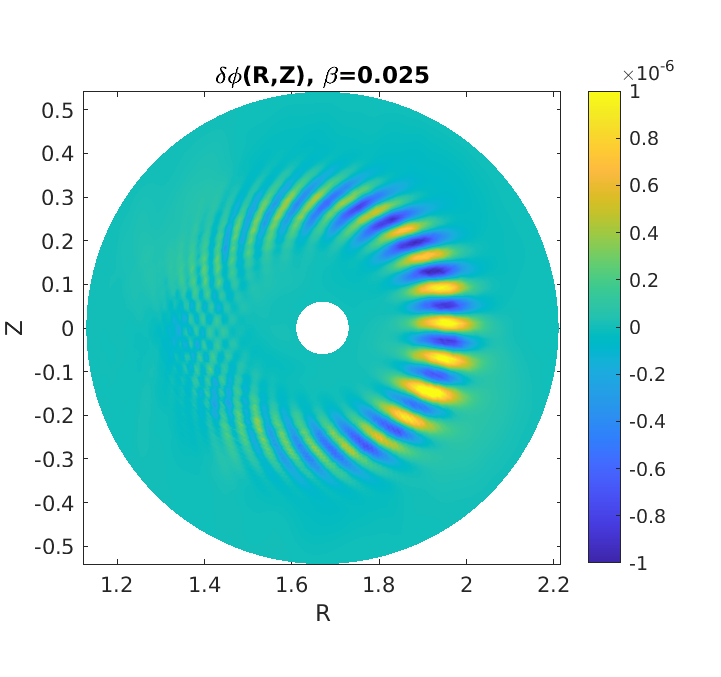}
\includegraphics[width=0.4\textwidth]{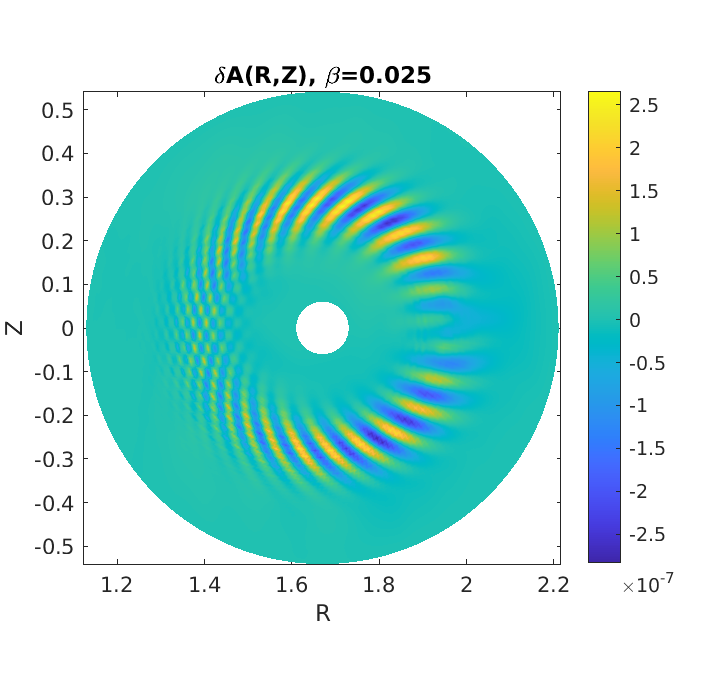}
\caption{The 2D mode structures of $\delta\phi$ and $\delta A_\|$ at different values of $\beta$ for the Cyclone case in Section \ref{subsec:cbc_itgkbm}. Upper frames: ITG mode for $\beta=0.001$.  $|\delta\Phi/\delta{A}_\| |\gtrsim 50$. Lower frames: KBM for $\beta=0.025$.  $|\delta\Phi/\delta{A}_\|  |\approx 4$.}
\label{fig:mode2d_cbc}
\end{figure}

\subsection{Electrostatic ion temperature gradient mode in TCV}
\label{subsec:tcv}
The TCV-X21 case has been studied  experimentally and numerically for Tokamak à configuration variable (TCV) \cite{oliveira2022validation,body2022development,ulbl2023influence}. This case is used by various codes such as GRILLIX and GENE-X for the studies of the transport and the profile generation with the consideration of the separatrix. The core plasma simulation is performed in this section to demonstrate the basic features of the linear ITG instability. The electrostatic model is adopted for single-$n$ simulations. The reference Larmor radius is $\rho_N=0.39012$ cm. The reference magnetic field is $B_{\rm{ref}}=0.90727$ T. 
The ion-to-electron mass ratio used is 100. 
The mode structures in the linear and nonlinear stages are shown in Fig.~\ref{fig:mode2d_tcv} for the $n=10$ harmonic. The nonlinear mode structure is broadened in the radial direction, consistent with our previous observations \cite{lu2019development,lu2025piecewise}. 
The time evolution of the $n=10$ harmonic is shown in Fig.~\ref{fig:growth_tcv}. The mode is saturated due to the wave-particle nonlinear interaction. We take the data in the exponentially growing stage to calculate the growth rate. 
The growth rate of different $n$ harmonics is shown in the right frame of Fig.~\ref{fig:growth_tcv}. For the simulations without finite Larmor radius effect, the growth rate increases as the toroidal mode number $n$ increases, until $n\le20$, where the finite orbit effect becomes significant. With the finite Larmor radius effect, the ITG growth rates are the largest for $n\in[10,15]$. 

\begin{figure}[t]
\centering
\includegraphics[width=0.48\textwidth]{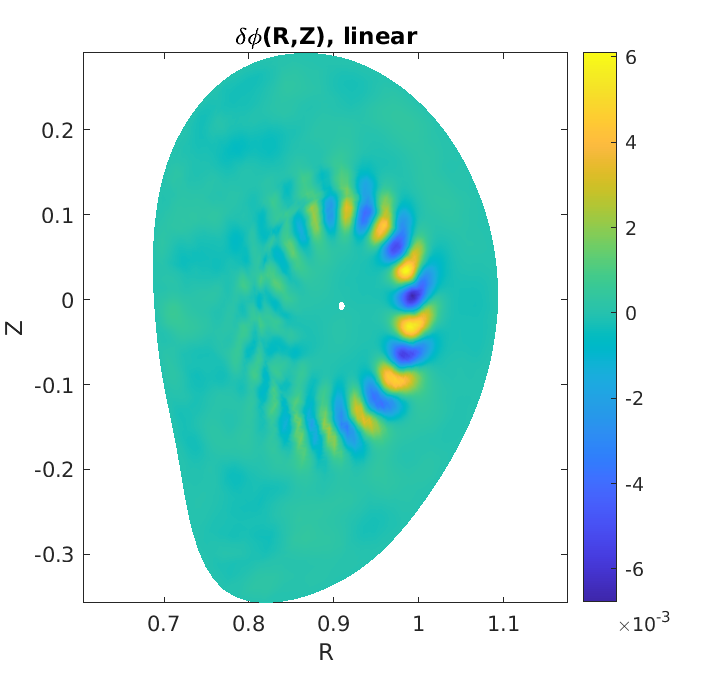}
\includegraphics[width=0.48\textwidth]{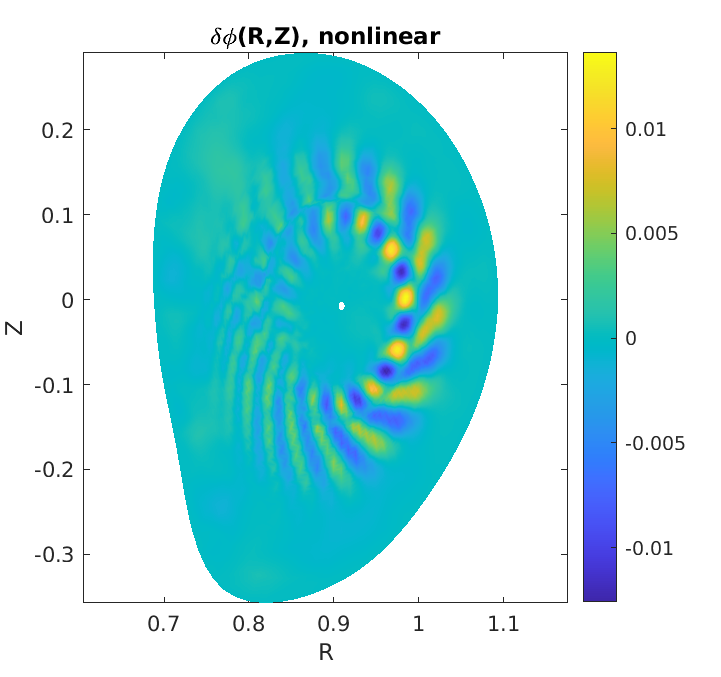}
\caption{Linear and early-nonlinear-stage mode structure of the $n=10$ ITG mode in the TCV tokamak plasma (Section \ref{subsec:tcv}).}
\label{fig:mode2d_tcv}
\end{figure}

\label{subsec:tcv}
\begin{figure}[t]
\centering
\includegraphics[width=0.48\textwidth]{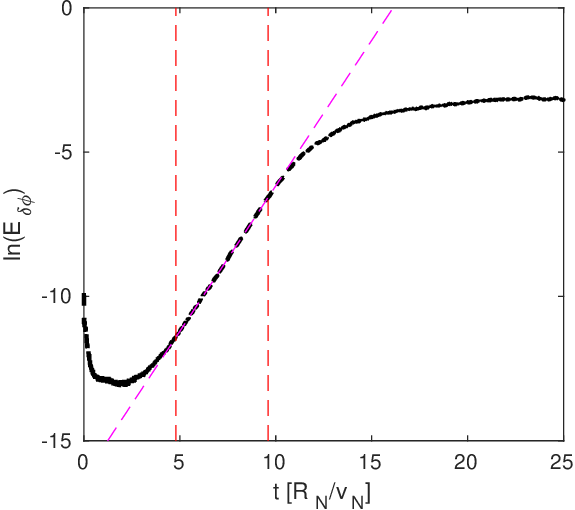}
\includegraphics[width=0.48\textwidth]{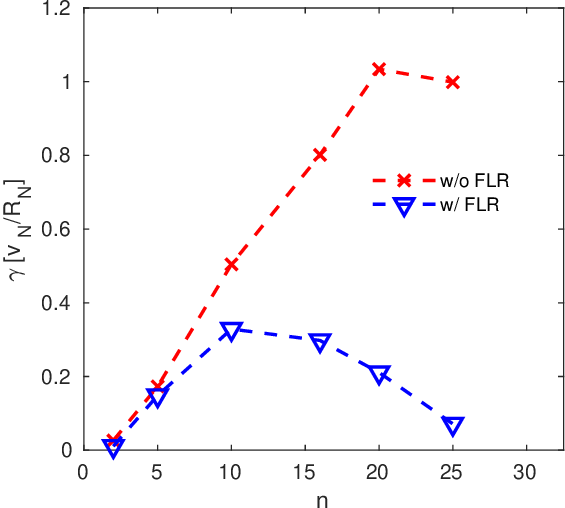}
\caption{Left: the time evolution of the $n=10$ simulation, where the vertical dashed lines indicate the data used to calculate the growth rate, and the dashed magenta line denotes the fitted time evolution constructed from the fitted growth rate. Right: growth rate of the ITG mode versus toroidal mode number $n$ with and without finite Larmor radius effect of thermal ions (Section \ref{subsec:tcv}).}
\label{fig:growth_tcv}
\end{figure}

\subsection{Electromagnetic simulations of ITG mode/KBM with a $(n,\beta)$ dependency}
\label{subsec:kbm_gknet}
The economical Cyclone case is adopted for the multi-$n$ simulation of the ITG mode and KBM using the 2D1F solver and the 3D piecewise field-aligned finite element method. By using the ion-to-electron mass ratio $m_{\rm i}/m_{\rm e}=100$ instead of $1836$, the time-step size $\Delta t$ can be increased by $\sim\sqrt{18.36}\approx4.285$ times. In addition, by using $\rho^*\equiv\rho_{\rm i}/a=100$ instead of $180$, the computational cost is reduced by  a factor of $1/1.8^3\approx1/5.832$. The total computational cost is reduced by a factor of $\sim 1/25$. 

The $n=10$ toroidal harmonic is simulated for different values of $\beta$. The economic Cyclone case has also been studied by the GKNET code \cite{ishizawa2019global}. As demonstrated in the left frame of Fig.~\ref{fig:growth beta005020}, the GKX results agree with the GKNET results except for a minor difference at the ITG-KBM transition region ($1.2\%<\beta<1.5\%$). The difference can be due to the different treatment in the equilibrium and higher-order terms in the gyro-average. Nevertheless, the agreement is reasonably good for most values of $\beta$.  

Two sets of simulations are run for $\beta=0.5\%, 2\%$ respectively. 
The single-$n$ simulations are run using the 2D1F solver to identify the linear growth rate for different values of $n$. For each single-$n$ simulation, we use $1\times10^6$ ions with a $4$ point gyro average. Most cases are run using $16\times10^6$ electrons, except for the $n\ge18$ cases $32\times10^6$ electrons are needed. For $\beta=0.5\%$, the maximum growth rate is at $n\approx12$, while for $\beta=2\%$, the maximum growth rate is at $n\approx7$ as shown in the right frame of Fig.~\ref{fig:growth beta005020}. As $\beta$ increases, the toroidal mode number $n$ of the most unstable toroidal harmonic downshifts and the maximum growth rate increases, as shown in the figure, which is consistent with the previous theoretical analyses \cite{aleynikova2017quantitative}. 
The 3D simulation is run on 32 nodes (AMD Genoa EPYC 9354) of the TOK cluster at the Max Planck Institute for Plasma Physics, with 32 CPU cores on each node,  with the processor base frequency of $3.25$ GHz and a max turbo frequency of $3.8$~GHz. The multi-$n$ simulation of the electrostatic ITG mode can be found in our previous work \cite{lu2025piecewise}. In this work, we only demonstrate the multi-$n$ simulation of the KBM at $\beta=2\%$. 
The nonlinear simulations start with pure noise. In total of $64\times10^6$ electron markers and $10^6$ ion markers are simulated. The time-step size is $\Delta t=0.005t_N$. For each $R_N/v_N$, it takes $\sim1.5$ hours. 
The $n$ spectrum is measured at the end of the exponential growing stage, as shown in Fig.~\ref{fig:energy1d_multin_gknet}. The peak of the $n$ spectrum at $n=7$ is consistent with that of the linear growth rate. 
The most unstable $n$ component grows the fastest, and thus the growth rate calculated using the total field energy is close to that of the most unstable mode. 
The mode structure within the $r/a=0.5$ flux surface is plotted versus the poloidal and toroidal coordinates in Fig.~\ref{fig:mode2d_pfafem_gknet}. The field-aligned structure is captured by the piecewise field-aligned finite element solver, as shown in the left frame. The field value in a denser grid in $(\theta,\phi)$ is interpolated, and the Fourier component $\delta\phi_{m,n}(r)$ is shown in the right frame. The magnitude peaks near $nq=m$ and the dominant harmonics are near $n=7$, which is consistent with the $n$ spectrum shown in Fig.~\ref{fig:energy1d_multin_gknet}. While in this work, we focus on the numerical studies of the GKX code in multi-$n$ simulations, more physics studies will be reported in a separate work.

\begin{figure}[t]
\centering
\includegraphics[width=0.48\textwidth]{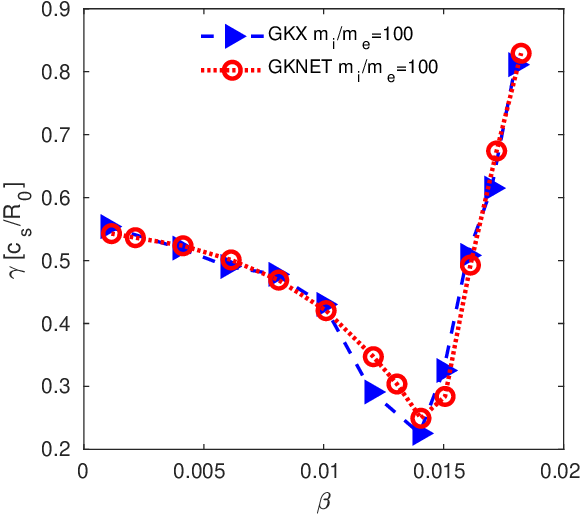}
\includegraphics[width=0.48\textwidth]{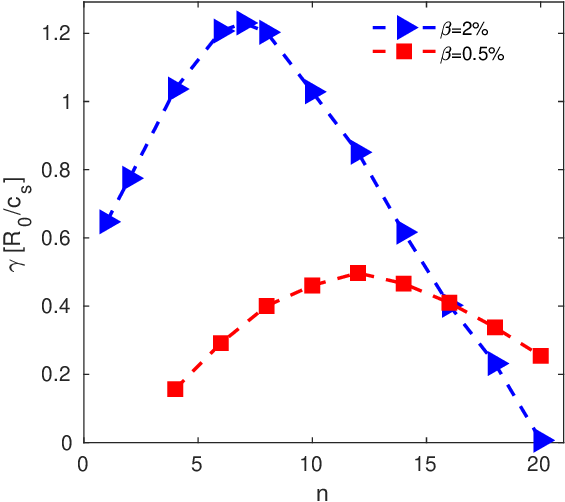}
\caption{Left: growth rate of the ITG mode/KBM versus $\beta$ and the comparison with GKNET results using the economic Cyclone parameters in Section \ref{subsec:kbm_gknet}. 
Right: growth rate versus the toroidal mode number $n$ for $\beta=0.5\%, 2\%$. }
\label{fig:growth beta005020}
\end{figure}

\begin{figure}[t]
\centering
\includegraphics[width=0.48\textwidth]{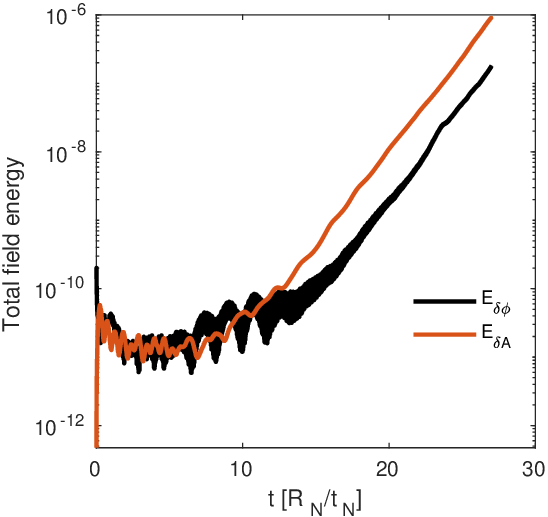}
\includegraphics[width=0.48\textwidth]{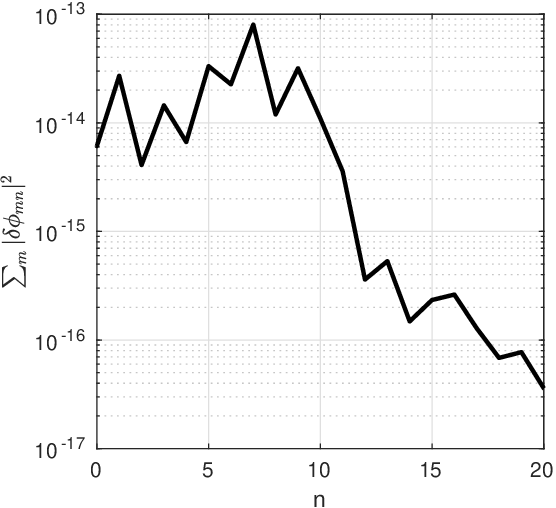}
\caption{The time evolution of the total field energy for $\beta=2\%$ (left) and the $n$ spectrum in the linear stage (right) of the multiple harmonic simulation of KBM in Section \ref{subsec:kbm_gknet}. }
\label{fig:energy1d_multin_gknet}
\end{figure}

\begin{figure}[t]
\centering
\includegraphics[width=0.98\textwidth]{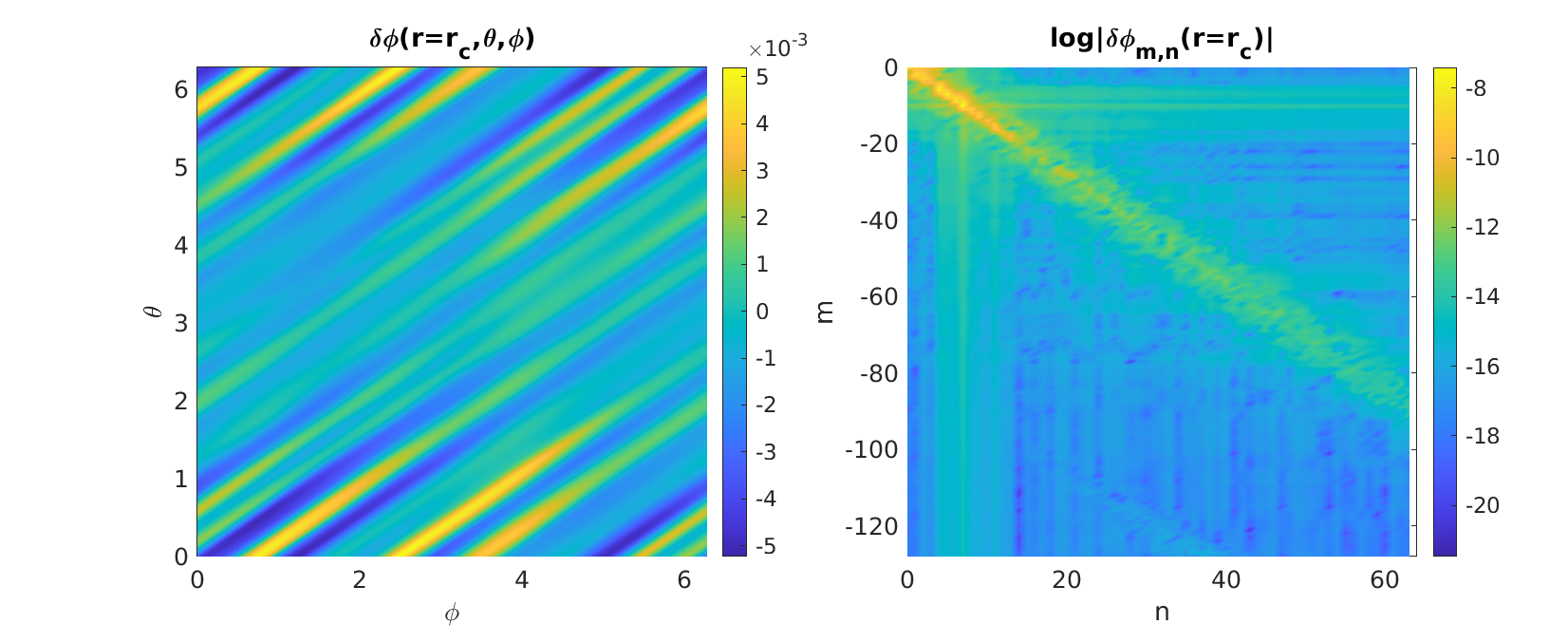}
\caption{The linear mode structure versus the poloidal and toroidal coordinates in the flux surface at $r/a=0.5$ (left) and the Fourier components (right).}
\label{fig:mode2d_pfafem_gknet}
\end{figure}

\section{Conclusions}
\label{sec:conclusion}
 The TRIMEG-GKX code has integrated the electromagnetic features, field-aligned finite elements, multi-species capabilities, rigorous strong form in the gyrocenter equations of motion, high flexibility of single or multiple harmonic simulations, and cache optimization as a unique combination among other gyrokinetic codes.  These features make it a powerful tool for investigating turbulence and energetic particle (EP) dynamics in core plasmas. 
This work summarizes the recent developments in the TRIMEG-GKX code, highlighting the underlying physics models, numerical methods, and representative simulation results. Key features and advancements of the code include:

\begin{enumerate} 
    \item Application of object-oriented programming, which facilitates flexible and efficient implementation of multi-species simulations. 
    \item Development of a piecewise field-aligned finite element method to enhance grid efficiency along the magnetic field direction, significantly improving the performance of multi-$n$ simulations. 
    \item Implementation of cache optimization techniques to increase memory access efficiency for interpolating perturbed field data at particle locations, achieving a tenfold speed-up for typical single-harmonic simulations without requiring particle sorting. 
    \item A rigorous treatment of the strong form in the electromagnetic gyrocenter equations of motion, improving accuracy and stability of the pullback-mixed variable scheme, thereby enabling simulations without relying on Fourier filters and numerical buffers. 
\end{enumerate}

Numerical benchmarks confirm the capabilities of the TRIMEG-GKX code to simulate a wide range of micro- and macro-instabilities, including Alfvén eigenmodes driven by energetic particles, ion temperature gradient (ITG) modes, and kinetic ballooning modes. These have been demonstrated in both ad hoc and experimentally reconstructed equilibria, such as those from ASDEX Upgrade (AUG), Tokamak à configuration variable (TCV), and the Joint European Torus (JET). Future work will focus on the study of nonlinear physics, along with systematic efforts in validation and verification. Planned developments include the treatment of open field line configurations and collisions, building on our prior work \cite{lu2025high}, and in conjunction with the development of the TRIMEG-C1 code \cite{lu2024gyrokinetic}.

\section*{ACKNOWLEDGEMENTS} 
Z.X. Lu appreciates the fruitful discussions with Alexey Mishchenko, Peiyou Jiang, Ralf Kleiber, Matthias Borchardt, Eric Sonnendrücker, and Fulvio Zonca. The simulations in this work were run on the TOK cluster and the MPCDF Viper/Raven supercomputers.  The Eurofusion projects TSVV-8, ACH/MPG and ATEP are acknowledged. This work has been carried out within the framework of the EUROfusion Consortium, funded by the European Union via the Euratom Research and Training Programme (Grant Agreement No 101052200—EUROfusion). Views and opinions expressed are however those of the author(s) only and do not necessarily reflect those of the European Union or the European Commission. Neither the European Union nor the European Commission can be held responsible for them.


\appendix
\section{Gyrocenter equation of motion in $(r,\phi,\theta)$}
\label{gcmotion}
The parallel velocity and the magnetic drift velocity in equilibrium are given by  
\begin{eqnarray}
\dot{r}_0&=&b_0^{*r}v_\|+C_d\frac{F}{J}\partial_\theta B\;\;,\\
\dot\theta_0&=&b_0^{*\theta}v_\| -C_d\frac{F}{J}\partial_r B  \;\;,\\
\dot\phi_0&=&b_0^{*\phi}v_\| +C_d\frac{\partial_r\psi}{R^2} (g^{rr}\partial_rB+g^{r\theta }\partial_\theta B)\;\;, \\
C_d &=& \frac{\bar{m}_s}{\bar{e}_s}\rho_N\frac{B_{\rm ref}}{BB_\|^*}\mu\;\;,
\end{eqnarray}
where $b_0^\alpha\equiv{\bb}_0\cdot\nabla\alpha$ is the contravariant component of $\bb_0$,  $g^{\alpha\beta}\equiv\nabla\alpha\cdot\nabla\beta$ is the metric tensor.  
The leading order terms are as follows 
\begin{eqnarray}
\dot{r}_0&=&C_d\frac{F}{J}\partial_\theta B\;\;,\\
\dot\theta_0&=&\frac{B^\theta}{B}v_\| -C_d\frac{F}{J}\partial_r B  \;\;,\\
\dot\phi_0&=&\frac{B^\phi}{B}v_\| +C_d\frac{\partial_r\psi}{R^2} (g^{rr}\partial_rB+g^{r\theta }\partial_\theta B)\;\;, \\
C_d &=& \frac{\bar{m}_s}{\bar{e}_s}\rho_N\frac{B_{\rm ref}}{B^3}(v_\|^2+\mu B)\;\;,
\end{eqnarray}
where $B^\alpha\equiv{\bB}\cdot\nabla\alpha$ is the contravariant component of the equilibrium magnetic field, and the magnetic curvature is replaced as an approximation with the gradient in the magnetic drift term.

The equation due to the mirror force is given by
\begin{eqnarray}
\dot{v}_{\|,0}=-\frac{\mu}{JB}\partial_r\psi\partial_\theta B\;\;.
\end{eqnarray}

Regarding the perturbed equations of motion, the equilibrium variables are calculated in $(r,\phi,\theta)$ coordinates but $\partial_r\langle\delta\Phi\rangle$, $\partial_\theta\langle\delta\Phi\rangle$ and $\partial_\phi\langle\delta\Phi\rangle$ are calculated in $r,\phi,\eta$ coordinates. Especially, ${\bb}\cdot\nabla\langle\delta\Phi\rangle$ is directly calculated in $(r,\phi,\eta)$ using ${\bb}\cdot\nabla={\bb}\cdot\nabla\phi\partial/\partial\phi|_{r,\eta}$.

The $E\times B$ velocity is given by 
\begin{eqnarray}\label{eq:r1dot}
\dot{r}_1 &=& -C_{E,\theta}g^{rr}\partial_\phi\langle\delta{G}\rangle 
+C_{E,\phi}\partial_\theta \langle\delta{G}\rangle
-\frac{\bar e_s}{\bar m_s} \langle\delta A^h_{\|} \rangle b^{*,r} \;\;, \\
\dot\theta_1 &=& -C_{E,\theta}g^{r\theta} \partial_\phi\langle\delta{G}\rangle-C_{E,\phi}\partial_r\langle\delta{G}\rangle
-\frac{\bar e_s}{\bar m_s} \langle\delta A^h_{\|} \rangle b^{*,\theta} \;\;,\\
\dot\phi_1&=&C_{E,\theta}
\left(g^{rr}\partial_r\langle\delta{G}\rangle+g^{r\theta}\partial_\theta\langle\delta{G}\rangle\right)\
-\frac{\bar e_s}{\bar m_s} \langle\delta A^h_{\|} \rangle b^{*,\phi} \;\;, \\
C_{E,\theta}&=&\rho_{\rm ref}B_{\rm ref}\frac{\partial_r\psi}{B^2R^2} \;\;,\;\;
C_{E,\phi} = \rho_{\rm ref}B_{\rm ref} \frac{F}{JB^2} \;\;,
\end{eqnarray}
where $\delta{G}=\delta\Phi-v_\|\delta{A}_\|$, $\partial_r\langle\delta{G}\rangle$, $\partial_\theta\langle\delta{G}\rangle$ and $\partial_\phi\langle\delta{G}\rangle$ are in $(r,\phi,\theta)$ coordinates.  If the 3D field-aligned FEM solver is adopted, $\partial_r\delta G$, $\partial_\theta\delta G$ and $\partial_\phi\delta G$ need to be calculated from the Clebsch coordinates $r,\phi,\eta$. Using the chain rule, we readily have
\begin{eqnarray}
\partial_r|_{\phi,\theta} &=&\partial_r|_{\phi,\eta} +(\partial_r\eta)\partial_\eta\;\;,  \\
\partial_\theta|_{r,\phi} &=& (\partial_\theta\eta)\partial_\eta  \;\;, \\
\partial_\phi|_{r,\theta} &=&\partial_\phi|_{r,\eta} +(\partial_\phi\eta)\partial_\eta  \;\;.
\end{eqnarray}
The gyro-average $\langle\delta\Phi\rangle$ is calculated in $(r,\theta,\phi)$ coordinates.  
The contribution from the parallel perturbed field is 
\begin{eqnarray}
    -\frac{e_s}{m_s}{\bf b}\cdot\nabla\langle\delta\Phi\rangle=-\frac{e_s}{m_s}\frac{B^\phi}{B}\partial_\phi|_{r,\eta}\langle\delta\Phi\rangle\;\;.
\end{eqnarray}





\begin{thebibliography}{40}
  \expandafter\ifx\csname natexlab\endcsname\relax\def\natexlab#1{#1}\fi
  \providecommand{\url}[1]{\texttt{#1}}
  \providecommand{\href}[2]{#2}
  \providecommand{\path}[1]{#1}
  \providecommand{\DOIprefix}{doi:}
  \providecommand{\ArXivprefix}{arXiv:}
  \providecommand{\URLprefix}{URL: }
  \providecommand{\Pubmedprefix}{pmid:}
  \providecommand{\doi}[1]{\href{http://dx.doi.org/#1}{\path{#1}}}
  \providecommand{\Pubmed}[1]{\href{pmid:#1}{\path{#1}}}
  \providecommand{\bibinfo}[2]{#2}
  \ifx\xfnm\relax \def\xfnm[#1]{\unskip,\space#1}\fi
  \bibitem[{Lee(1983)}]{lee1983gyrokinetic}
  \bibinfo{author}{W.~W. Lee},
  \newblock \bibinfo{title}{Gyrokinetic approach in particle simulation},
  \newblock \bibinfo{journal}{Phys. Fluids} \bibinfo{volume}{26}
    (\bibinfo{year}{1983}) \bibinfo{pages}{556}.
  \bibitem[{Lin et~al.(1998)Lin, Hahm, Lee, Tang, and White}]{lin1998turbulent}
  \bibinfo{author}{Z.~Lin}, \bibinfo{author}{T.~S. Hahm}, \bibinfo{author}{W.~W.
    Lee}, \bibinfo{author}{W.~M. Tang}, \bibinfo{author}{R.~B. White},
  \newblock \bibinfo{title}{Turbulent transport reduction by zonal flows:
    Massively parallel simulations},
  \newblock \bibinfo{journal}{Science} \bibinfo{volume}{281}
    (\bibinfo{year}{1998}) \bibinfo{pages}{1835--1837}.
  \bibitem[{Chang et~al.(2017)Chang, Ku, Tynan, Hager, Churchill, Cziegler,
    Greenwald, Hubbard, and Hughes}]{chang2017fast}
  \bibinfo{author}{C.~Chang}, \bibinfo{author}{S.~Ku},
    \bibinfo{author}{G.~Tynan}, \bibinfo{author}{R.~Hager},
    \bibinfo{author}{R.~Churchill}, \bibinfo{author}{I.~Cziegler},
    \bibinfo{author}{M.~Greenwald}, \bibinfo{author}{A.~Hubbard},
    \bibinfo{author}{J.~Hughes},
  \newblock \bibinfo{title}{Fast low-to-high confinement mode bifurcation
    dynamics in a tokamak edge plasma gyrokinetic simulation},
  \newblock \bibinfo{journal}{Phys. Rev. Lett.} \bibinfo{volume}{118}
    (\bibinfo{year}{2017}) \bibinfo{pages}{175001}.
  \bibitem[{Hatzky et~al.(2019)Hatzky, Kleiber, K{\"o}nies, Mishchenko,
    Borchardt, Bottino, and Sonnendr{\"u}cker}]{hatzky2019reduction}
  \bibinfo{author}{R.~Hatzky}, \bibinfo{author}{R.~Kleiber},
    \bibinfo{author}{A.~K{\"o}nies}, \bibinfo{author}{A.~Mishchenko},
    \bibinfo{author}{M.~Borchardt}, \bibinfo{author}{A.~Bottino},
    \bibinfo{author}{E.~Sonnendr{\"u}cker},
  \newblock \bibinfo{title}{Reduction of the statistical error in electromagnetic
    gyrokinetic particle-in-cell simulations},
  \newblock \bibinfo{journal}{Journal of Plasma Physics} \bibinfo{volume}{85}
    (\bibinfo{year}{2019}).
  \bibitem[{Chen and Parker(2007)}]{chen2007electromagnetic}
  \bibinfo{author}{Y.~Chen}, \bibinfo{author}{S.~E. Parker},
  \newblock \bibinfo{title}{Electromagnetic gyrokinetic $\delta$f
    particle-in-cell turbulence simulation with realistic equilibrium profiles
    and geometry},
  \newblock \bibinfo{journal}{Journal of Computational Physics}
    \bibinfo{volume}{220} (\bibinfo{year}{2007}) \bibinfo{pages}{839--855}.
  \bibitem[{Mishchenko et~al.(2019)Mishchenko, Bottino, Biancalani, Hatzky,
    Hayward-Schneider, Ohana, Lanti, Brunner, Villard, Borchardt
    et~al.}]{mishchenko2019pullback}
  \bibinfo{author}{A.~Mishchenko}, \bibinfo{author}{A.~Bottino},
    \bibinfo{author}{A.~Biancalani}, \bibinfo{author}{R.~Hatzky},
    \bibinfo{author}{T.~Hayward-Schneider}, \bibinfo{author}{N.~Ohana},
    \bibinfo{author}{E.~Lanti}, \bibinfo{author}{S.~Brunner},
    \bibinfo{author}{L.~Villard}, \bibinfo{author}{M.~Borchardt}, et~al.,
  \newblock \bibinfo{title}{Pullback scheme implementation in {ORB5}},
  \newblock \bibinfo{journal}{Computer Physics Communications}
    \bibinfo{volume}{238} (\bibinfo{year}{2019}) \bibinfo{pages}{194--202}.
  \bibitem[{Lu et~al.(2021)Lu, Meng, Hoelzl, and Lauber}]{lu2021development}
  \bibinfo{author}{Z.~X. Lu}, \bibinfo{author}{G.~Meng},
    \bibinfo{author}{M.~Hoelzl}, \bibinfo{author}{P.~Lauber},
  \newblock \bibinfo{title}{The development of an implicit full f method for
    electromagnetic particle simulations of {A}lfv{\'e}n waves and energetic
    particle physics},
  \newblock \bibinfo{journal}{Journal of Computational Physics}
    \bibinfo{volume}{440} (\bibinfo{year}{2021}) \bibinfo{pages}{110384}.
  \bibitem[{Sturdevant et~al.(2021)Sturdevant, Ku, Chac{\'o}n, Chen, Hatch, Cole,
    Sharma, Adams, Chang, Parker et~al.}]{sturdevant2021verification}
  \bibinfo{author}{B.~J. Sturdevant}, \bibinfo{author}{S.~Ku},
    \bibinfo{author}{L.~Chac{\'o}n}, \bibinfo{author}{Y.~Chen},
    \bibinfo{author}{D.~Hatch}, \bibinfo{author}{M.~Cole},
    \bibinfo{author}{A.~Sharma}, \bibinfo{author}{M.~Adams},
    \bibinfo{author}{C.~Chang}, \bibinfo{author}{S.~Parker}, et~al.,
  \newblock \bibinfo{title}{Verification of a fully implicit particle-in-cell
    method for the $v_{||}$-formalism of electromagnetic gyrokinetics in the xgc
    code},
  \newblock \bibinfo{journal}{Physics of Plasmas} \bibinfo{volume}{28}
    (\bibinfo{year}{2021}) \bibinfo{pages}{072505}.
  \bibitem[{Chen et~al.(2021)Chen, Chen, Zonca, and Lin}]{chen2021gyrokinetic}
  \bibinfo{author}{L.~Chen}, \bibinfo{author}{H.~Chen},
    \bibinfo{author}{F.~Zonca}, \bibinfo{author}{Y.~Lin},
  \newblock \bibinfo{title}{A gyrokinetic simulation model for low frequency
    electromagnetic fluctuations in magnetized plasmas},
  \newblock \bibinfo{journal}{Science China Physics, Mechanics \& Astronomy}
    \bibinfo{volume}{64} (\bibinfo{year}{2021}) \bibinfo{pages}{245211}.
  \bibitem[{Rosen et~al.(2022)Rosen, Lu, and Hoelzl}]{rosen2022and}
  \bibinfo{author}{M.~Rosen}, \bibinfo{author}{Z.~Lu},
    \bibinfo{author}{M.~Hoelzl},
  \newblock \bibinfo{title}{An e and b gyrokinetic simulation model for kinetic
    {A}lfv{\'e}n waves in tokamak plasmas},
  \newblock \bibinfo{journal}{Physics of Plasmas} \bibinfo{volume}{29}
    (\bibinfo{year}{2022}).
  \bibitem[{Bao et~al.(2018)Bao, Lin, and Lu}]{bao2018conservative}
  \bibinfo{author}{J.~Bao}, \bibinfo{author}{Z.~Lin}, \bibinfo{author}{Z.~Lu},
  \newblock \bibinfo{title}{A conservative scheme for electromagnetic simulation
    of magnetized plasmas with kinetic electrons},
  \newblock \bibinfo{journal}{Physics of Plasmas} \bibinfo{volume}{25}
    (\bibinfo{year}{2018}).
  \bibitem[{Lanti et~al.(2020)Lanti, Ohana, Tronko, Hayward-Schneider, Bottino,
    McMillan, Mishchenko, Scheinberg, Biancalani, Angelino
    et~al.}]{lanti2020orb5}
  \bibinfo{author}{E.~Lanti}, \bibinfo{author}{N.~Ohana},
    \bibinfo{author}{N.~Tronko}, \bibinfo{author}{T.~Hayward-Schneider},
    \bibinfo{author}{A.~Bottino}, \bibinfo{author}{B.~F. McMillan},
    \bibinfo{author}{A.~Mishchenko}, \bibinfo{author}{A.~Scheinberg},
    \bibinfo{author}{A.~Biancalani}, \bibinfo{author}{P.~Angelino}, et~al.,
  \newblock \bibinfo{title}{{ORB5}: A global electromagnetic gyrokinetic code
    using the {PIC} approach in toroidal geometry},
  \newblock \bibinfo{journal}{Computer Physics Communications}
    \bibinfo{volume}{251} (\bibinfo{year}{2020}).
  \bibitem[{Kleiber et~al.(2024)Kleiber, Borchardt, Hatzky, K{\"o}nies, Leyh,
    Mishchenko, Riemann, Slaby, Garc{\'\i}a-Rega{\~n}a, Sanchez
    et~al.}]{kleiber2024euterpe}
  \bibinfo{author}{R.~Kleiber}, \bibinfo{author}{M.~Borchardt},
    \bibinfo{author}{R.~Hatzky}, \bibinfo{author}{A.~K{\"o}nies},
    \bibinfo{author}{H.~Leyh}, \bibinfo{author}{A.~Mishchenko},
    \bibinfo{author}{J.~Riemann}, \bibinfo{author}{C.~Slaby},
    \bibinfo{author}{J.~Garc{\'\i}a-Rega{\~n}a}, \bibinfo{author}{E.~Sanchez},
    et~al.,
  \newblock \bibinfo{title}{{EUTERPE}: A global gyrokinetic code for stellarator
    geometry},
  \newblock \bibinfo{journal}{Computer Physics Communications}
    \bibinfo{volume}{295} (\bibinfo{year}{2024}) \bibinfo{pages}{109013}.
  \bibitem[{Taimourzadeh et~al.(2019)Taimourzadeh, Bass, Chen, Collins,
    Gorelenkov, K{\"o}nies, Lu, Spong, Todo, Austin
    et~al.}]{taimourzadeh2019verification}
  \bibinfo{author}{S.~Taimourzadeh}, \bibinfo{author}{E.~Bass},
    \bibinfo{author}{Y.~Chen}, \bibinfo{author}{C.~Collins},
    \bibinfo{author}{N.~Gorelenkov}, \bibinfo{author}{A.~K{\"o}nies},
    \bibinfo{author}{Z.~Lu}, \bibinfo{author}{D.~A. Spong},
    \bibinfo{author}{Y.~Todo}, \bibinfo{author}{M.~Austin}, et~al.,
  \newblock \bibinfo{title}{Verification and validation of integrated simulation
    of energetic particles in fusion plasmas},
  \newblock \bibinfo{journal}{Nuclear Fusion} \bibinfo{volume}{59}
    (\bibinfo{year}{2019}) \bibinfo{pages}{066006}.
  \bibitem[{Kraus et~al.(2017)Kraus, Kormann, Morrison, and
    Sonnendr{\"u}cker}]{kraus2017gempic}
  \bibinfo{author}{M.~Kraus}, \bibinfo{author}{K.~Kormann},
    \bibinfo{author}{P.~J. Morrison}, \bibinfo{author}{E.~Sonnendr{\"u}cker},
  \newblock \bibinfo{title}{{GEMPIC}: geometric electromagnetic particle-in-cell
    methods},
  \newblock \bibinfo{journal}{Journal of Plasma Physics} \bibinfo{volume}{83}
    (\bibinfo{year}{2017}) \bibinfo{pages}{905830401}.
  \bibitem[{Meng et~al.(2025)Meng, Kormann, Poulsen, and
    Sonnendruecker}]{meng2025geometric}
  \bibinfo{author}{G.~Meng}, \bibinfo{author}{K.~Kormann},
    \bibinfo{author}{E.~Poulsen}, \bibinfo{author}{E.~Sonnendruecker},
  \newblock \bibinfo{title}{A geometric particle-in-cell discretization of the
    drift-kinetic and fully kinetic vlasov-maxwell equations},
  \newblock \bibinfo{journal}{Plasma Physics and Controlled Fusion}
    (\bibinfo{year}{2025}).
  \bibitem[{Lu et~al.(2023)Lu, Meng, Hatzky, Hoelzl, and Lauber}]{lu2023full}
  \bibinfo{author}{Z.~X. Lu}, \bibinfo{author}{G.~Meng},
    \bibinfo{author}{R.~Hatzky}, \bibinfo{author}{M.~Hoelzl},
    \bibinfo{author}{P.~Lauber},
  \newblock \bibinfo{title}{Full f and $\delta$f gyrokinetic particle simulations
    of {Alfv\'en} waves and energetic particle physics},
  \newblock \bibinfo{journal}{Plasma Physics and Controlled Fusion}
    (\bibinfo{year}{2023}).
  \bibitem[{Lu et~al.(2025)Lu, Meng, Sonnendr{\"u}cker, Hatzky, Mishchenko,
    Zonca, Lauber, and Hoelzl}]{lu2025piecewise}
  \bibinfo{author}{Z.~X. Lu}, \bibinfo{author}{G.~Meng},
    \bibinfo{author}{E.~Sonnendr{\"u}cker}, \bibinfo{author}{R.~Hatzky},
    \bibinfo{author}{A.~Mishchenko}, \bibinfo{author}{F.~Zonca},
    \bibinfo{author}{P.~Lauber}, \bibinfo{author}{M.~Hoelzl},
  \newblock \bibinfo{title}{Piecewise field-aligned finite element method for
    multi-mode nonlinear particle simulations in tokamak plasmas},
  \newblock \bibinfo{journal}{Journal of Plasma Physics} \bibinfo{volume}{91}
    (\bibinfo{year}{2025}) \bibinfo{pages}{E48}.
  \bibitem[{Meng et~al.(2020)Meng, Lauber, Lu, and Wang}]{meng2020effects}
  \bibinfo{author}{G.~Meng}, \bibinfo{author}{P.~Lauber}, \bibinfo{author}{Z.~X.
    Lu}, \bibinfo{author}{X.~Wang},
  \newblock \bibinfo{title}{Effects of the non-perturbative mode structure on
    energetic particle transport},
  \newblock \bibinfo{journal}{Nuclear Fusion} \bibinfo{volume}{60}
    (\bibinfo{year}{2020}) \bibinfo{pages}{056017}.
  \bibitem[{Lu et~al.(2019)Lu, Lauber, Hayward-Schneider, Bottino, and
    Hoelzl}]{lu2019development}
  \bibinfo{author}{Z.~X. Lu}, \bibinfo{author}{P.~Lauber},
    \bibinfo{author}{T.~Hayward-Schneider}, \bibinfo{author}{A.~Bottino},
    \bibinfo{author}{M.~Hoelzl},
  \newblock \bibinfo{title}{Development and testing of an unstructured mesh
    method for whole plasma gyrokinetic simulations in realistic tokamak
    geometry},
  \newblock \bibinfo{journal}{Phys. Plasmas} \bibinfo{volume}{26}
    (\bibinfo{year}{2019}) \bibinfo{pages}{122503}.
  \bibitem[{Rekhviashvili et~al.(2023)Rekhviashvili, Lu, Hoelzl, Bergmann, and
    Lauber}]{lana2023neoclassical}
  \bibinfo{author}{L.~Rekhviashvili}, \bibinfo{author}{Z.~Lu},
    \bibinfo{author}{M.~Hoelzl}, \bibinfo{author}{A.~Bergmann},
    \bibinfo{author}{P.~Lauber},
  \newblock \bibinfo{title}{Gyrokinetic simulations of neoclassical electron
    transport and bootstrap current generation in tokamak plasmas in the trimeg
    code},
  \newblock \bibinfo{journal}{Physics of Plasmas} \bibinfo{volume}{30}
    (\bibinfo{year}{2023}).
  \bibitem[{Lu et~al.(2024)Lu, Meng, Hatzky, Sonnendr{\"u}cker, Mishchenko,
    Lauber, Zonca, and Hoelzl}]{lu2024gyrokinetic}
  \bibinfo{author}{Z.~X. Lu}, \bibinfo{author}{G.~Meng},
    \bibinfo{author}{R.~Hatzky}, \bibinfo{author}{E.~Sonnendr{\"u}cker},
    \bibinfo{author}{A.~Mishchenko}, \bibinfo{author}{P.~Lauber},
    \bibinfo{author}{F.~Zonca}, \bibinfo{author}{M.~Hoelzl},
  \newblock \bibinfo{title}{Gyrokinetic electromagnetic particle simulations in
    triangular meshes with {C1} finite elements},
  \newblock \bibinfo{journal}{Plasma Physics and Controlled Fusion}
    \bibinfo{volume}{67} (\bibinfo{year}{2024}) \bibinfo{pages}{015015}.
  \bibitem[{Mishchenko et~al.(2014)Mishchenko, K{\"o}nies, Kleiber, and
    Cole}]{mishchenko2014pullback}
  \bibinfo{author}{A.~Mishchenko}, \bibinfo{author}{A.~K{\"o}nies},
    \bibinfo{author}{R.~Kleiber}, \bibinfo{author}{M.~Cole},
  \newblock \bibinfo{title}{Pullback transformation in gyrokinetic
    electromagnetic simulations},
  \newblock \bibinfo{journal}{Physics of Plasmas} \bibinfo{volume}{21}
    (\bibinfo{year}{2014}).
  \bibitem[{Mishchenko et~al.(2023)Mishchenko, Borchardt, Hatzky, Kleiber,
    K{\"o}nies, N{\"u}hrenberg, Xanthopoulos, Roberg-Clark, and
    Plunk}]{mishchenko2023global}
  \bibinfo{author}{A.~Mishchenko}, \bibinfo{author}{M.~Borchardt},
    \bibinfo{author}{R.~Hatzky}, \bibinfo{author}{R.~Kleiber},
    \bibinfo{author}{A.~K{\"o}nies}, \bibinfo{author}{C.~N{\"u}hrenberg},
    \bibinfo{author}{P.~Xanthopoulos}, \bibinfo{author}{G.~Roberg-Clark},
    \bibinfo{author}{G.~G. Plunk},
  \newblock \bibinfo{title}{Global gyrokinetic simulations of electromagnetic
    turbulence in stellarator plasmas},
  \newblock \bibinfo{journal}{Journal of Plasma Physics} \bibinfo{volume}{89}
    (\bibinfo{year}{2023}) \bibinfo{pages}{955890304}.
  \bibitem[{Williams(2008)}]{williams}
  \bibinfo{author}{J.~Williams}, \bibinfo{title}{{Bspline-Fortran}},
    \bibinfo{year}{2008}.
    \bibinfo{note}{{https://github.com/jacobwilliams/bspline-fortran}}.
  \bibitem[{Lu et~al.(2012)Lu, Zonca, and Cardinali}]{lu2012theoretical}
  \bibinfo{author}{Z.~X. Lu}, \bibinfo{author}{F.~Zonca},
    \bibinfo{author}{A.~Cardinali},
  \newblock \bibinfo{title}{Theoretical and numerical studies of wave-packet
    propagation in tokamak plasmas},
  \newblock \bibinfo{journal}{Physics of Plasmas} \bibinfo{volume}{19}
    (\bibinfo{year}{2012}).
  \bibitem[{Sauter and Medvedev(2013)}]{sauter2013cocos}
  \bibinfo{author}{O.~Sauter}, \bibinfo{author}{S.~Y. Medvedev},
  \newblock \bibinfo{title}{Tokamak coordinate conventions: {COCOS}},
  \newblock \bibinfo{journal}{Computer Physics Communications}
    \bibinfo{volume}{184} (\bibinfo{year}{2013}) \bibinfo{pages}{293--302}.
  \bibitem[{L{\"u}tjens et~al.(1996)L{\"u}tjens, Bondeson, and
    Sauter}]{lutjens1996chease}
  \bibinfo{author}{H.~L{\"u}tjens}, \bibinfo{author}{A.~Bondeson},
    \bibinfo{author}{O.~Sauter},
  \newblock \bibinfo{title}{The chease code for toroidal {MHD} equilibria},
  \newblock \bibinfo{journal}{Computer physics communications}
    \bibinfo{volume}{97} (\bibinfo{year}{1996}) \bibinfo{pages}{219--260}.
  \bibitem[{Meng et~al.(2022)Meng, Lauber, Wang, and Lu}]{guo2022mode}
  \bibinfo{author}{G.~Meng}, \bibinfo{author}{P.~Lauber},
    \bibinfo{author}{X.~Wang}, \bibinfo{author}{Z.~Lu},
  \newblock \bibinfo{title}{Mode structure symmetry breaking of reversed shear
    {A}lfv{\'e}n eigenmodes and its impact on the generation of parallel velocity
    asymmetries in energetic particle distribution},
  \newblock \bibinfo{journal}{Plasma Science and Technology} \bibinfo{volume}{24}
    (\bibinfo{year}{2022}) \bibinfo{pages}{025101}.
  \bibitem[{Garcia et~al.(2024)Garcia, Kazakov, Coelho, Dreval, de~la Luna,
    Solano, {\v{S}}tancar, Varela, Baruzzo, Belli et~al.}]{garcia2024stable}
  \bibinfo{author}{J.~Garcia}, \bibinfo{author}{Y.~Kazakov},
    \bibinfo{author}{R.~Coelho}, \bibinfo{author}{M.~Dreval},
    \bibinfo{author}{E.~de~la Luna}, \bibinfo{author}{E.~R. Solano},
    \bibinfo{author}{{\v{Z}}.~{\v{S}}tancar}, \bibinfo{author}{J.~Varela},
    \bibinfo{author}{M.~Baruzzo}, \bibinfo{author}{E.~Belli}, et~al.,
  \newblock \bibinfo{title}{Stable deuterium-tritium plasmas with improved
    confinement in the presence of energetic-ion instabilities},
  \newblock \bibinfo{journal}{Nature Communications} \bibinfo{volume}{15}
    (\bibinfo{year}{2024}) \bibinfo{pages}{7846}.
  \bibitem[{K{\"o}nies et~al.(2018)K{\"o}nies, Briguglio, Gorelenkov, Feh{\'e}r,
    Isaev, Lauber, Mishchenko, Spong, Todo, Cooper et~al.}]{konies2018benchmark}
  \bibinfo{author}{A.~K{\"o}nies}, \bibinfo{author}{S.~Briguglio},
    \bibinfo{author}{N.~Gorelenkov}, \bibinfo{author}{T.~Feh{\'e}r},
    \bibinfo{author}{M.~Isaev}, \bibinfo{author}{P.~Lauber},
    \bibinfo{author}{A.~Mishchenko}, \bibinfo{author}{D.~A. Spong},
    \bibinfo{author}{Y.~Todo}, \bibinfo{author}{W.~A. Cooper}, et~al.,
  \newblock \bibinfo{title}{Benchmark of gyrokinetic, kinetic {MHD} and gyrofluid
    codes for the linear calculation of fast particle driven tae dynamics},
  \newblock \bibinfo{journal}{Nucl. Fusion} \bibinfo{volume}{58}
    (\bibinfo{year}{2018}) \bibinfo{pages}{126027}.
  \bibitem[{Lauber et~al.(2018)Lauber, Geiger, Papp, Guimarais, Poloskei,
    Igochine, Maraschek, Pokol, Hayward-Schneider, Lu
    et~al.}]{lauber2018strongly}
  \bibinfo{author}{P.~Lauber}, \bibinfo{author}{B.~Geiger},
    \bibinfo{author}{G.~Papp}, \bibinfo{author}{L.~Guimarais},
    \bibinfo{author}{P.~Z. Poloskei}, \bibinfo{author}{V.~Igochine},
    \bibinfo{author}{M.~Maraschek}, \bibinfo{author}{G.~Pokol},
    \bibinfo{author}{T.~Hayward-Schneider}, \bibinfo{author}{Z.~Lu}, et~al.,
  \newblock \bibinfo{title}{Strongly non-linear energetic particle dynamics in
    {ASDEX Upgrade}scenarios with core impurity accumulation},
  \newblock \bibinfo{journal}{Proceedings of the 27th IAEA Fusion energy}
    (\bibinfo{year}{2018}).
  \bibitem[{Vlad et~al.(2021)Vlad, Wang, Vannini, Briguglio, Carlevaro, Falessi,
    Fogaccia, Fusco, Zonca, Biancalani et~al.}]{vlad2021linear}
  \bibinfo{author}{G.~Vlad}, \bibinfo{author}{X.~Wang},
    \bibinfo{author}{F.~Vannini}, \bibinfo{author}{S.~Briguglio},
    \bibinfo{author}{N.~Carlevaro}, \bibinfo{author}{M.~Falessi},
    \bibinfo{author}{G.~Fogaccia}, \bibinfo{author}{V.~Fusco},
    \bibinfo{author}{F.~Zonca}, \bibinfo{author}{A.~Biancalani}, et~al.,
  \newblock \bibinfo{title}{A linear benchmark between {HYMAGYC}, {MEGA} and
    {ORB5} codes using the {NLED-AUG} test case to study {A}lfv{\'e}nic modes
    driven by energetic particles},
  \newblock \bibinfo{journal}{Nuclear Fusion} \bibinfo{volume}{61}
    (\bibinfo{year}{2021}) \bibinfo{pages}{116026}.
  \bibitem[{G{\"o}rler et~al.(2016)G{\"o}rler, Tronko, Hornsby, Bottino, Kleiber,
    Norscini, Grandgirard, Jenko, and Sonnendr{\"u}cker}]{gorler2016intercode}
  \bibinfo{author}{T.~G{\"o}rler}, \bibinfo{author}{N.~Tronko},
    \bibinfo{author}{W.~A. Hornsby}, \bibinfo{author}{A.~Bottino},
    \bibinfo{author}{R.~Kleiber}, \bibinfo{author}{C.~Norscini},
    \bibinfo{author}{V.~Grandgirard}, \bibinfo{author}{F.~Jenko},
    \bibinfo{author}{E.~Sonnendr{\"u}cker},
  \newblock \bibinfo{title}{Intercode comparison of gyrokinetic global
    electromagnetic modes},
  \newblock \bibinfo{journal}{Phys. Plasmas} \bibinfo{volume}{23}
    (\bibinfo{year}{2016}).
  \bibitem[{Ishizawa et~al.(2019)Ishizawa, Imadera, Nakamura, and
    Kishimoto}]{ishizawa2019global}
  \bibinfo{author}{A.~Ishizawa}, \bibinfo{author}{K.~Imadera},
    \bibinfo{author}{Y.~Nakamura}, \bibinfo{author}{Y.~Kishimoto},
  \newblock \bibinfo{title}{Global gyrokinetic simulation of turbulence driven by
    kinetic ballooning mode},
  \newblock \bibinfo{journal}{Physics of Plasmas} \bibinfo{volume}{26}
    (\bibinfo{year}{2019}).
  \bibitem[{Oliveira et~al.(2022)Oliveira, Body, Galassi, Theiler, Laribi,
    Tamain, Stegmeir, Giacomin, Zholobenko, Ricci
    et~al.}]{oliveira2022validation}
  \bibinfo{author}{D.~Oliveira}, \bibinfo{author}{T.~Body},
    \bibinfo{author}{D.~Galassi}, \bibinfo{author}{C.~Theiler},
    \bibinfo{author}{E.~Laribi}, \bibinfo{author}{P.~Tamain},
    \bibinfo{author}{A.~Stegmeir}, \bibinfo{author}{M.~Giacomin},
    \bibinfo{author}{W.~Zholobenko}, \bibinfo{author}{P.~Ricci}, et~al.,
  \newblock \bibinfo{title}{Validation of edge turbulence codes against the
    {TCV-X21} diverted l-mode reference case},
  \newblock \bibinfo{journal}{Nuclear Fusion} \bibinfo{volume}{62}
    (\bibinfo{year}{2022}) \bibinfo{pages}{096001}.
  \bibitem[{Body(2022)}]{body2022development}
  \bibinfo{author}{T.~A.~J. Body}, \bibinfo{title}{Development of Turbulence
    Simulations for the Edge \& Divertor and Validation against Experiment},
    Ph.D. thesis, Technische Universit{\"a}t M{\"u}nchen, \bibinfo{year}{2022}.
  \bibitem[{Ulbl et~al.(2023)Ulbl, Body, Zholobenko, Stegmeir, Pfennig, and
    Jenko}]{ulbl2023influence}
  \bibinfo{author}{P.~Ulbl}, \bibinfo{author}{T.~Body},
    \bibinfo{author}{W.~Zholobenko}, \bibinfo{author}{A.~Stegmeir},
    \bibinfo{author}{J.~Pfennig}, \bibinfo{author}{F.~Jenko},
  \newblock \bibinfo{title}{Influence of collisions on the validation of global
    gyrokinetic simulations in the edge and scrape-off layer of {TCV}},
  \newblock \bibinfo{journal}{Physics of Plasmas} \bibinfo{volume}{30}
    (\bibinfo{year}{2023}).
  \bibitem[{Aleynikova and Zocco(2017)}]{aleynikova2017quantitative}
  \bibinfo{author}{K.~Aleynikova}, \bibinfo{author}{A.~Zocco},
  \newblock \bibinfo{title}{Quantitative study of kinetic ballooning mode theory
    in simple geometry},
  \newblock \bibinfo{journal}{Physics of Plasmas} \bibinfo{volume}{24}
    (\bibinfo{year}{2017}).
  \bibitem[{Lu et~al.(2025)Lu, Meng, Tyranowski, and Chankin}]{lu2025high}
  \bibinfo{author}{Z.~X. Lu}, \bibinfo{author}{G.~Meng},
    \bibinfo{author}{T.~Tyranowski}, \bibinfo{author}{A.~Chankin},
  \newblock \bibinfo{title}{High-order stochastic integration schemes for the
    {Rosenbluth-Trubnikov} collision operator in particle simulations},
  \newblock \bibinfo{journal}{Journal of Computational Physics}
    (\bibinfo{year}{2025}) \bibinfo{pages}{113811}.
  
  \end{thebibliography}
\end{document}